\documentclass[iop]{emulateapj}
\usepackage{epstopdf}

\newcommand{\ms}{\mbox{m s$^{-1}$}}
\newcommand{\msy}{\mbox{m s$^{-1}$ yr$^{-1}$}}
\newcommand{\kms}{\mbox{km s$^{-1}$}}
\newcommand{\msun}{M$_{\odot}$}
\newcommand{\rsun}{R$_{\odot}$}
\newcommand{\mjup}{M$_{\rm Jup}$}
\newcommand{\mearth}{M$_{\oplus}$}

\newcommand{\msini}{$M \sin i$}

\newcommand{\ab}{$\sim$}

\newcommand{\simgt}{\lower.5ex\hbox{$\; \buildrel > \over \sim \;$}}
\newcommand{\simlt}{\lower.5ex\hbox{$\; \buildrel < \over \sim \;$}}
\newcommand{\chisq}{$\chi_{\rm{\nu}}^2$}
\newcommand{\pr}{\rm{Pr}}

\shortauthors{Choi {\it et~al.\/}}
\shorttitle{Barnard's Star Planet Search}

\begin{document}
\title{Precise Doppler Monitoring of Barnard's Star\footnote[\dag]{B{\lowercase{ased on observations made at}} K{\lowercase{eck}} O{\lowercase{bservatory and}} L{\lowercase{ick}} O{\lowercase{bservatory}}}}

\author{Jieun Choi\altaffilmark{1},
Chris McCarthy\altaffilmark{2},
Geoffrey W. Marcy\altaffilmark{1,2},
Andrew W. Howard\altaffilmark{1},
Debra A. Fischer\altaffilmark{3},
John A. Johnson\altaffilmark{4},
Howard Isaacson\altaffilmark{1},
Jason T. Wright\altaffilmark{5,6}}

\email{jieun\_eb@berkeley.edu}

\altaffiltext{1}{Department of Astronomy, University of California,
Berkeley, CA 94720, USA}

\altaffiltext{2}{Department of Physics and Astronomy,
San Francisco State University, San Francisco, CA 94132, USA}

\altaffiltext{3}{Department of Astronomy, Yale University, New Haven, CT 06520-8101, USA}

\altaffiltext{4}{Department of Astronomy, California Institute of Technology, Pasadena, CA 91125, USA}

\altaffiltext{5}{Department of Astronomy, The Pennsylvania State University, University Park, PA 16802, USA}

\altaffiltext{6}{Center for Exoplanets and Habitable Worlds, The Pennsylvania State University, University Park, PA 16802, USA}

\begin{abstract}
We present 248 precise Doppler measurements of Barnard's Star (Gl 699), the second nearest star system to Earth, obtained from Lick and Keck Observatories during 25 years between 1987 and 2012. The early precision was 20 \ms{} but was 2 \ms{} during the last 8 years, constituting the most extensive and sensitive search for Doppler signatures of planets around this stellar neighbor. We carefully analyze the 136 Keck radial velocities spanning 8 years by first applying a periodogram analysis to search for nearly circular orbits. We find no significant periodic Doppler signals with amplitudes above $\sim$2 \ms{}, setting firm upper limits on the minimum mass (\msini) of any planets with orbital periods from 0.1 to 1000 days. Using a Monte Carlo analysis for circular orbits, we determine that planetary companions to Barnard's Star with masses above 2 \mearth{} and periods below 10 days would have been detected. Planets with periods up to 2 years and masses above 10 \mearth{} (0.0
3 \mjup) are also ruled out. A similar analysis allowing for eccentric orbits yields comparable mass limits. The habitable zone of Barnard's Star appears to be devoid of roughly Earth-mass planets or larger, save for face-on orbits. Previous claims of planets around the star by van de Kamp are strongly refuted. The radial velocity of Barnard's Star increases with time at $4.515\pm0.002$ \msy{}, consistent with the predicted geometrical effect, secular acceleration, that exchanges transverse for radial components of velocity.
\end{abstract}

\section{Introduction} \label{intro}

To date, over 700 exoplanets have been identified orbiting other stars \citep{Marcy2008,Mayor2011, Wright2011}, and another 2300 exoplanet candidates have been found from the Kepler spaceborne telescope \citep{Batalha2012}, the majority of which are real planets \citep{Morton2011, Lissauer2012}. Hundreds of exoplanets have now been discovered within 50 pc,
most by precision Doppler surveys \citep{Wright2011}. These nearest
exoplanets provide the best opportunities for follow-up observations
by the next generation of planet detection techniques, which now
include numerous strategies, both ground- and space-based, such as direct imaging \citep{Marois2008}, transit (e.g., \citealt{Irwin2009, Muirhead2012b, Berta2012}), IR thermal signatures (e.g., \citealt{Charbonneau2005}), and astrometry \citep{Anglada2012}.

Searching the nearest stars for planets presents special challenges. These campaigns require large telescopes to conduct exhaustive long-term radial velocity (RV) surveys, and the very closest stars---those within a few pc---are mostly faint M dwarfs. While nearly 300 M dwarfs are currently
being monitored for exoplanets \citep{Johnson2010b, Delfosse2012},
relatively little radial velocity data on them were available until
recently. The first planet orbiting an M dwarf was discovered in 2001
around Gl 876 \citep{Marcy2001}. This M4V star has since been found to
host four companions, including a 7.5 Earth-mass planet \citep{Rivera2005, Rivera2010}. In the last few years, many planets have been found around other M dwarfs, including: 
Gl 832 (M1.5, \citealt{Bailey2009}),
Gl 649 (M2, \citealt{Johnson2010b}),
Gl 179 (M3.5, \citealt{Howard2010}),
HIP 12961 (M0, \citealt{Forveille2011}),
Gl 676 A (M0, \citealt{Forveille2011})
Gl 433 (M1.5 \citealt{Bonfils2011}), and
Gl 667 C (M1, \citealt{Bonfils2011, Delfosse2012}),
increasing the number of currently known planetary companions around M dwarfs to 25 \citep{Wright2011} at the time of writing.

\cite{Johnson2007a} and \cite{Johnson2010c} found a positive correlation between the
frequency of jovian planets and host star mass, lending support to the
core accretion model of planet formation (e.g., \citealt{Kennedy2008}). It has been well-established that Jovian planets appear to form less frequently around M dwarfs than more massive stars \citep{Johnson2010b}. Currently, the best estimates for occurrence rate of planets with $M_{\rm P} \sin i > 0.3$ \mjup{} in orbits within 2.5 AU of their parent stars is $3.4^{+2.2}_{-0.9}\%$ for stars with $M_{\rm S} < 0.6$ \msun{}, where $M_{\rm P}$ and $M_{\rm S}$ refer to the masses of the planet and the star, respectively, compared to $\sim8\%$ for F,G, and K stars \citep{Cumming2008, Johnson2010b, Bonfils2011}. Recent works from both transit and RV surveys revealed that low-mass planets, rather than gas giants, are common around M dwarfs \citep{Bonfils2011, Howard2011}. Surveys with a long time baseline and high precision such as this work are necessary for the detection of these low-mass planets.

Nearby stars with high proper motions exhibit changes in their
RVs over time due to {\it secular acceleration} \citep{Stumpff1985}, an effect just at the limit of detectability for most surveys \citep{Kurster2003}. We remove the effect of secular acceleration from our RVs and search these data for signals due to exoplanets. We use a Monte Carlo approach to place upper limits on the minimum mass of possible exoplanets. Finally we compare these observations to previous claims of planetary companions around Barnard's Star.

\section{Properties of Barnard's Star}

At $1.824\pm0.005$ pc \citep{Cutri2003}, Barnard's Star
(Gl 699, HIP 87937, G 140-24, LHS 57) is the second
closest system and the fourth closest individual star to the Sun. It has been studied extensively since \cite{Barnard1916} discovered its nonpareil proper motion using first
epoch plates from Lick Observatory made in 1894. The properties of
Barnard's Star have been reviewed by \cite{Kurster2003}, \cite{Dawson2004}, and \cite{Paulson2006}. Its properties are summarized in Table\ 1.

Several factors suggest that the age of Barnard's Star exceeds 10
billion years. Its absolute radial velocity and total velocity with respect to the local standard of rest are 110 \kms \citep{Marcy1989} and 142 \kms \citep{Nidever2002}, respectively. This high space
motion suggests it is a halo star. It has a low metallicity, where [M/H] and [Fe/H] are found to be $-0.27\pm0.12$ and $-0.39\pm0.17$, respectively \citep{Muirhead2012a, RojasAyala2012, Muirhead2012b}. Additionally, \cite{Benedict1998} found ``very
weak evidence'' for photometric variability on a timescale of 130 days.
If this is interpreted as the rotation period of the star, then all of the above constitute further evidence for the star's advanced age \citep{Irwin2011}. 

The first planet search around Barnard's Star began 75 years ago at
Sproul observatory. In a series of papers, \cite{vandeKamp1963, vandeKamp1969a, vandeKamp1969blah, vandeKamp1975, vandeKamp1982} reported the detection of first one, then two, roughly Jupiter-mass companions based on multiple sets of astrometric data on photographic plates
obtained variously between 1912 and 1981. An independent astrometric
study by \cite{Gatewood1973} failed to confirm van de Kamp's results. New analysis of astrometric plates from McCormick Observatory \citep{Bartlett2006} did not reveal any significant perturbations. Space-based astrometric observations with the
HST Fine Guidance Sensor \citep{Benedict1999} and radial velocity
observations spanning 2.5 years \citep{Kurster2003} and 6 years \citep{Zechmeister2009} have also been reported, constraining the planet search space further. A comprehensive
review of astrometric and other planet searches around Barnard's Star
can be found in \cite{Bartlett2006} study of McCormick astrometric data on Barnard's Star. While concerns have been raised regarding the
possible systematic errors in the Sproul data \citep{Hershey1973, vandeKamp1982}, it is remarkable that almost half a century after the first exoplanet claims around Barnard's Star, no RV study in literature has tested van de Kamp's planetary system hypothesis.

\section{Radial Velocity Data}\label{rv_data_section}
We have observed Barnard's Star since 1987, at both Lick and Keck
observatories. In this span of time, our instrumental setup underwent
one significant improvement at each observatory. These ``fixes''
resulted in marked reduction in RV errors: after a
refurbishment at Lick in 1995, typical errors reduced from \ab 20 \ms{} to 10--15 \ms{}. Similarly, improvements in Keck hardware and
software in 2004 reduced errors from 3 \ms{} to \ab 1 \ms{} for most
stars \citep{Rivera2005}. For M dwarfs as faint as $V$ = 12, typical
errors are $\sim3$--5 \ms{}.

Our Keck data for Barnard's Star are affected by a one-time instrumental improvement, occurring in August 2004, with an upgraded detector with smaller pixels and improved charge-transfer efficiency. Since the upgrade, the RMS scatter has decreased from 4.2 \ms{} to 2.5 \ms{} on average, and the internal error has reduced from 1.9 \ms{} to 1.2 \ms{}. These errors are estimated as follows. Each observation is divided into $\sim700$ segments of $\sim2$ \AA{} in width, and the radial velocity is measured for each spectral segment \citep{Marcy2005, Johnson2007a}. The final RV reported is the weighted average of these velocity measurements and the corresponding internal error is the weighted uncertainty in the mean, which takes into account both photon-limited errors and wavelength-dependent errors that independently cause scatter in the measured velocities among the spectral segments \citep{Butler1996}.

While the aforementioned instrumental improvements are crucial to the ongoing success of any precise radial velocity (PRV) project, new detectors may introduce a one-time
velocity offset, and necessitate care in interpretation \citep{Crepp2012}. We compare the data obtained before the improvements (``pre-fix'') to those obtained after (``post-fix''). Since there is an absence of significant slopes in the pre- and post-fix data taken separately, we determine the value of the offset (4.2 \ms{}) simply by calculating the difference in the median of the pre- and post-fix data sets. By correcting for this constant offset, the older data can be combined with the new data, but a bias is also introduced: any sudden change in the long-term velocity of the star at this epoch would be suppressed. The pre- and post-fix data do not show any indication for such sudden velocity changes.

To ensure that the source of this velocity discontinuity is not astrophysical, we examine the Keck RVs of other M dwarfs to determine whether or not this feature is seen in other data sets as well. A sample of Keck M dwarfs is selected with the requirement that the semi-amplitude of Keplerian signals, $K$, is $\lesssim$ 10 \ms{} (no high amplitude signals due to the presence of planets, brown dwarfs, or stellar-mass companions, which could skew our calculations), which were observed both before and after August 2004. There are a total of 8 such objects in addition to Barnard's Star. We split the RVs of our entire sample to pre- and post-fix data sets and compute their median. To prevent the large number of data points of Barnard's Star---208 compared to an average of 50 data points each for other M dwarfs---from skewing the average, we calculate the velocity jump excluding Barnard's Star and obtain $2.0\pm0.7$ \ms{}. This velocity offset is added to pre-fix data and we procee
d with the adjusted velocities henceforth.

To search for the longest period RV signals, the (higher quality) Keck
and the (longer time baseline) Lick data are merged and calibrated to the same reference frame using an offset derived from contemporaneous data from both observatories. The difference in the median velocities of $3.4\pm2.5$ \ms{} is applied to the Lick data.

\subsection{Secular Acceleration}\label{sa}

The search for planetary systems akin to our own solar system is a
multi-decadal effort in which signals gradually emerge if precision
can be maintained over long timescales; the detection of a jovian companion to an M dwarf in a nine year orbit \citep{Bailey2009} is an excellent example. In RV data, a linear trend
may be the first sign of a subsequently well detected planet \citep{McCarthy2004}. However, as discussed by \cite{Stumpff1985}, \cite{vandeKamp1977b}, and \cite{vandeKamp1986}, secular acceleration will
also induce an approximately linear signal in differential radial
velocity measurements of fast moving stars. Secular acceleration is
purely a geometrical effect caused by the changing admixture of the
radial and transverse components of the velocity vector as the star
passes by the Sun.

To first order\footnote{The full
non-relativistic effect is given in \cite{Kurster2003} as Eq. 4. It
deviates from linear by less than 1\% even on timescales of many
decades. A relativistic correction to the observed proper motion \citep{Stumpff1985} causes a modification which is similarly negligible
for present purposes.} this effect is given by
\begin{equation}
{\rm{SA}} = \frac{0.0229 \mu^2}{\pi}\,\msy\;\;,
\end{equation}
where $\mu$ is the total proper motion in arcseconds per year and $\pi$
is the parallax in arcseconds \citep{vandeKamp1986}. The magnitude of
the effect implies that it can only be detected for the closest,
fastest stars in long-term precise velocity surveys. \cite{Kurster2003} reported a detection of secular
acceleration (SA) in 46 RVs of Barnard's Star spanning 2.5 years. They found a slope in the differential radial velocities of $2.97\pm0.51$ \msy{}, which is inconsistent with both constant velocity and with the
expected secular acceleration, $4.515\pm0.002$ \msy{}. However, they reported that if a third of their data is discarded, the resulting slope ($5.15\pm0.89$ \msy{}) is in agreement with the predicted value. They noted a correlation between observed RV and the
strength of the H-$\alpha$ line. \cite{Zechmeister2009} presented 29 more RVs for a total of 75 measurements spanning 6.5 years. In addition to the secular acceleration for which they adopted $4.497\pm0.012$ \msy{}, they measured a slope of $-0.688$ \msy{} in their data. 

\section{Analysis of Velocities of Barnard's Star}\label{analysis}
We obtained 40 observations of Barnard's Star from Lick Observatory
between 1987 and 2006, and 208 observations from Keck spanning 15 years. We correct for the motion of the observatory about the solar system barycenter by using the JPL ephemeris of the Solar System, JPLEPH.405 \citep{Standish1998}, giving the velocity vector of the Keck Observatory evaluated at the time of the photon-weighted midpoint of the exposure. We use the JPL ephemeris\footnote{http://ssd.jpl.nasa.gov/}, accessed with the IDL Astronomy User's Library\footnote{http://idlastro.gsfc.nasa.gov/} and our own driver
codes. We carried out extensive tests of our barycentric transformation code, finding discrepancies of 0.1 \ms{} in comparison with the analogous pulsar-timing code, TEMPO 1.1. We do not include the effects of the solar gravitational potential at the barycenter of the solar system (near the surface of the Sun) nor the gravitational blueshift caused by starlight falling into the potential well of the Sun at the location of the Earth, a $\sim3$ \ms{} effect. We also do not take into account the fixed gravitational redshift as light departs the photosphere of Barnard's star, an effect of hundreds of \ms{} that depends on its stellar mass and radius. We further ignore the convective blueshift of the starlight caused by the Doppler asymmetry between the upwelling hot gas and the downflowing cool gas. In addition to the solar system barycentric correction, we also remove a secular acceleration slope (Eq. 1) of $4.515\pm0.002$ \msy{} (accumulating to $\sim113$ \ms{} over 25 years). 
To compute the secular acceleration, we adopt a parallax of $0.5454\pm0.0003$ arcsec and a total proper motion of $10.3700\pm0.0003$ arcsec yr$^{-1}$ \citep{Benedict1999}. If H-$\alpha$ variability causes apparent fluctuations in the RVs of Barnard's Star \citep{Kurster2003}, then they are likely to average out over the long time baseline of years.

The RMS scatters of the total Keck data, pre-, and post-upgrade are 3.3 \ms{}, 4.2 \ms{}, and 2.5 \ms{}. The levels of variation are consistent with the internal errors, 1.4 \ms{}, combined with expected stellar jitter level of 2 \ms{} for M-dwarfs \citep{Johnson2007a} in quadrature, giving {\em{no indication of planetary companions}}. Stellar jitter accounts for uncertainties due to various activities on the star, such as surface convective motions, magnetic activity, rotation, and starspots \citep{Saar1997}. The entire 19 year combined Lick/Keck data set scatters by only 6.2 \ms{}. No significant slope or curvature in the data is found. 

\begin{figure}[!b]
\includegraphics[scale=0.8]{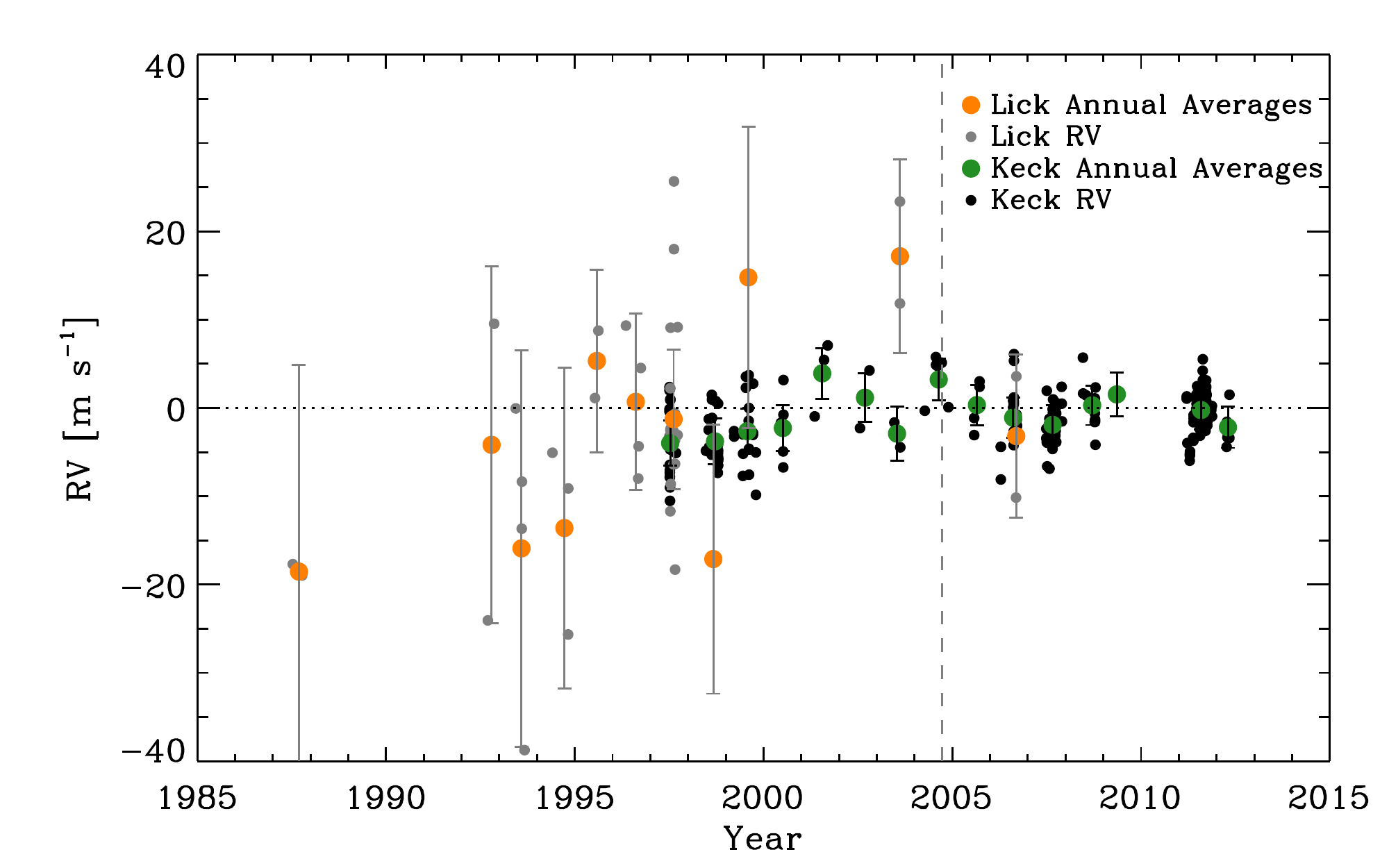}
\caption{RV measurements from Keck and Lick Observatories between 1987 and 2012. Gray and black dots denote individual unbinned RV measurements while orange and green circles are annual averages for Lick and Keck Observatories, respectively. Error bars are displayed for annual averages only. The vertical dashed line is shown to denote the time of Keck instrumental upgrade in August 2004. A jitter of 2 \ms{} is added in quadrature to the internal errors, and corrections for the velocity offsets between both Lick and Keck data and pre- and post-fix data are included.}
\label{lickkeck}
\end{figure}

We present RVs from Lick and Keck observatories for Barnard's Star in Fig.\ \ref{lickkeck}. Gray and black dots denote individual unbinned RV measurements while orange and green circles are annual averages for Lick and Keck Observatories, respectively. The error bars, corresponding to the standard error of the mean of individual measurements within a one year-bin, reflect both internal errors and jitter, which are added in quadrature. The vertical dashed line shows the time of Keck instrumental upgrade in August 2004. The RV measurements are listed in Tables\ 2 and 3 for Keck and Lick, respectively. The first column gives the UT date and the second column gives the BJD, the Barycentric Julian Date when the mid-point of the light train from the star would have crossed the barycenter of the solar system.  The third column gives the measured relative Doppler velocity reflecting the adjustments described in Section \ref{rv_data_section} and the fourth column gives the total uncertainty including both internal errors and jitter.

Since there is no indication of a long period signal from the combined 25 years of Lick plus Keck data, we scrutinize the more precise Keck data spanning only 15 years for Keplerian signals
with amplitudes near or below the typical error level. We use all RVs to search for long-term velocity trends but utilize only the higher-quality post-upgrade RVs (after 2004 August) for our analysis henceforth and bin the data by 2 hour intervals, unless noted otherwise.

\subsection{Circular Orbits}
We begin by restricting our analyses to circular orbits. First, we compute a Lomb-Scargle periodogram \citep{Lomb1976, Scargle1982} of the RVs, which shows the likelihood of the period of a trial sinusoid as a function of the period, similar to a Fourier power spectrum. Then we associate each peak with a False Alarm Probability (FAP) which depends on its amplitude and on the number of independent frequencies being searched. We search for the tallest peak in the periodogram, and for this candidate Keplerian signal, an FAP is computed as follows. We scramble the velocities while keeping the observation times fixed to produce a realistic estimate of the ``noise'' in the RVs from both instrumental and stellar sources. In fact, this method is overly conservative because any Doppler signals from planets buried in the RVs, when scrambled, will be adopted as extra noise. We compute a periodogram for 5000 such realizations, recording the amplitude of the tallest peak each time. The FA
P is the fraction of these noise trials that have larger amplitudes in the periodogram than the candidate peak. A false alarm probability can then be assigned to any periodicity.

A significant source of noise and aliasing arise due to the uneven and discrete sampling of data. For a real signal with power at some frequency $f_{\rm signal}$, our sampling results in power at other frequencies $f = f_{\rm signal} \pm n f_{\rm sampling}$, where $n$ is an integer. A {\it spectral window function} contains such signatures of our observation cadence, and by examining these features and comparing them to the periodogram of the data, we can untangle erroneous signal from real signal such as that produced by a planet.

\begin{figure}
\includegraphics[scale=0.8]{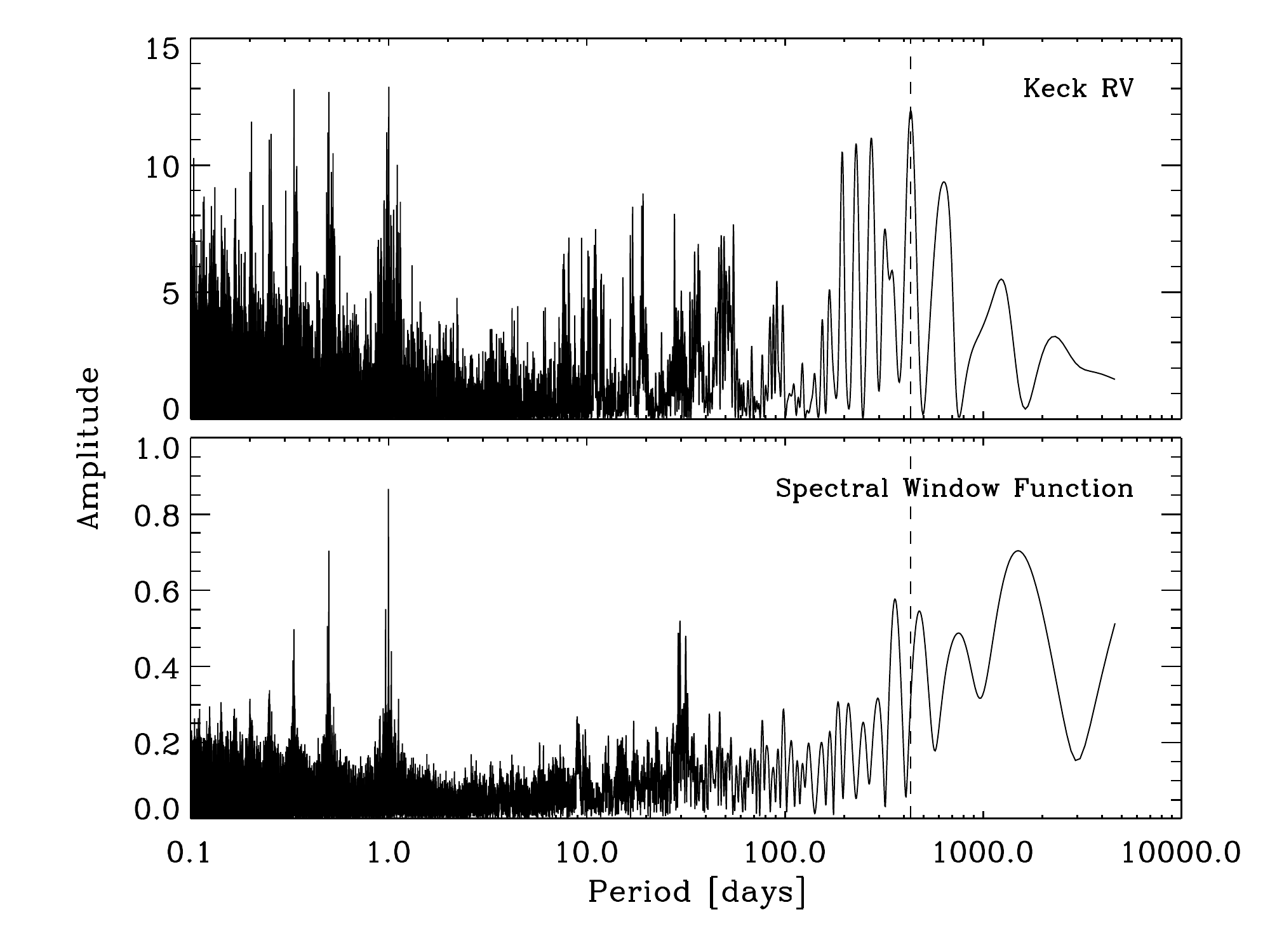}
\caption{Keck RV periodogram and the spectral window function (SWF). A dashed line is shown over the candidate signal at 430 days, but this periodicity is only marginally significant with a formal FAP of 1.84\%.}
\label{periodogram}
\end{figure}

To construct a spectral window function, $W$, as a function of frequency, we adopt the definition from \cite{Roberts1987}:

\begin{equation}
W(\nu) = \frac{1}{N}\sum_{r=0}^{N} e^{-2 \pi i \nu t_r}\;\;,
\end{equation}
where $t$ and $N$ are times and total number of observations, respectively. Any power we see will be solely due to the uneven and discrete sampling of the data. Fig.\ \ref{periodogram} shows the periodogram of the Keck RVs computed for periods ranging between 0.1 and 5000 days in the top panel and the spectral window function in the bottom panel.

Due to observation cadence, periodogram peaks occur at 1 sidereal year, 1 sidereal day, 1 solar day, and 1 synodic month \citep{Dawson2010}. These periodicities arise because times of observations are governed by the star's visibility in the sky throughout the year and day as well as the allocated telescope time near full moon. Our observations at the Keck telescope occurred only at night and usually during ``bright time" within a week of full moon. The spectral window function in Fig.\ \ref{periodogram} shows prominent peaks near 1.0 day, 29.5 days, and 365 days. The series of tall peaks below 1 day are higher harmonics of the 1 day alias.

\begin{figure}
\includegraphics[scale=0.8]{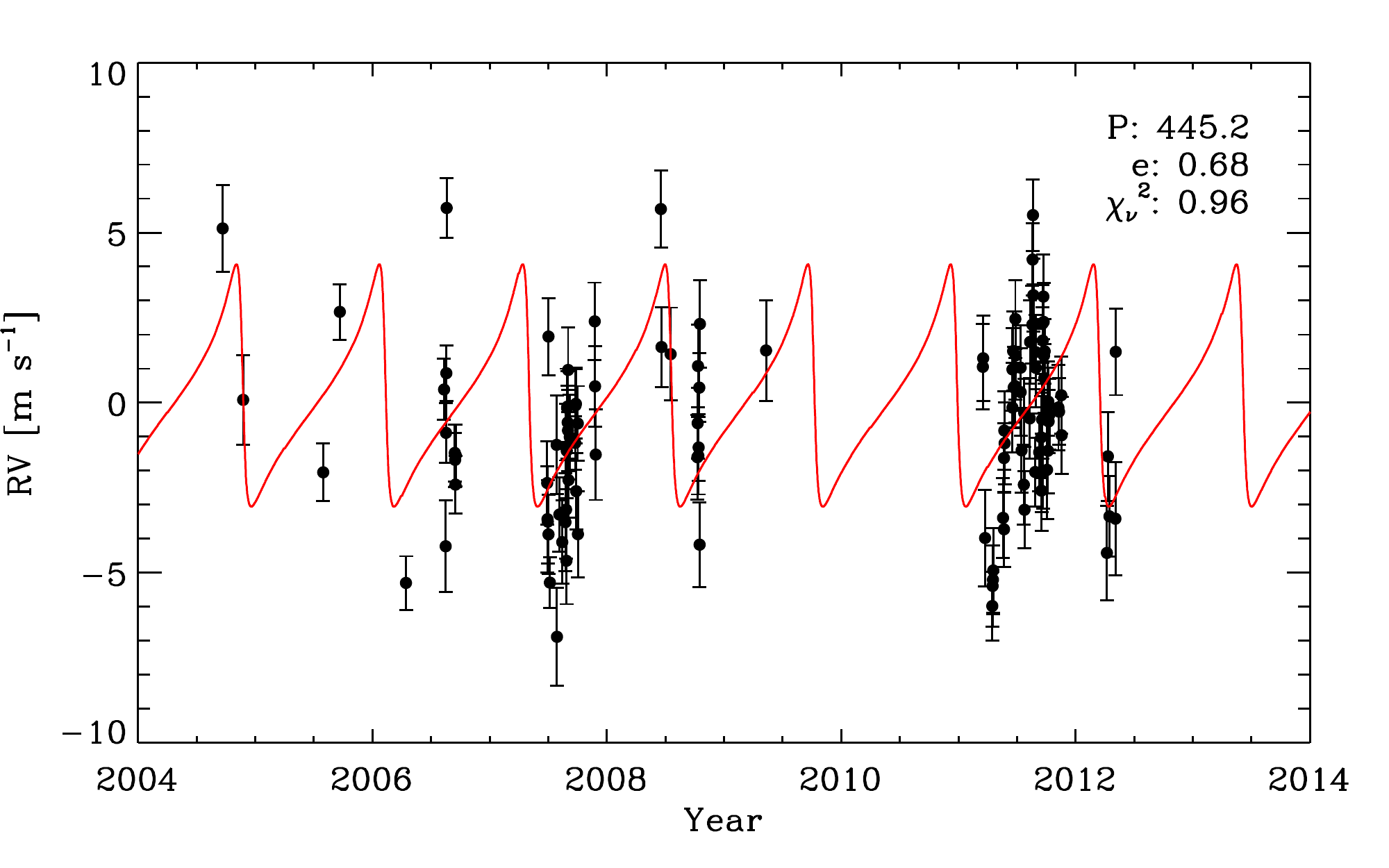}
\caption{Best Keplerian fit for a period of roughly 430 days and a maximum eccentricity of 0.8. The poor fit suggests that the interpretation of this periodicity as a Keplerian signal is dubious.}
\label{430d_fit}
\end{figure}

A weak signal found in the periodogram occurs at 430 days as shown by the dashed line in Fig.\ \ref{periodogram}. This peak has a formal FAP of 1.84\%, which is not convincing. We search for a best-fitting Keplerian orbit by allowing the period to float within a few days of 430 days, but orbital models yield unconvincing fits (see Fig.\ \ref{430d_fit}), casting further doubt on this 430 day periodicity.

\begin{figure}
\includegraphics[scale=0.8]{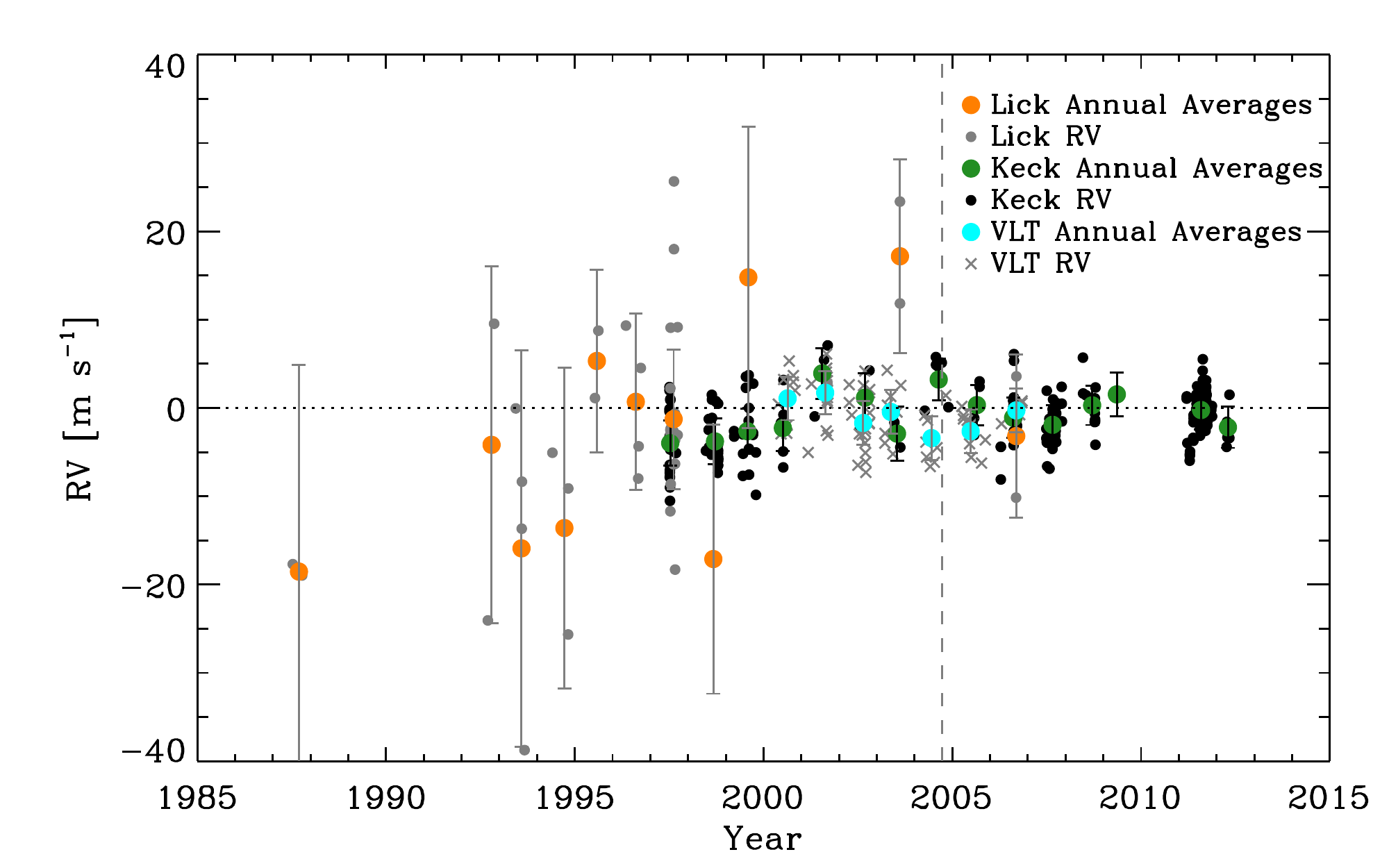}
\caption{RV measurements from Keck, Lick, and European Southern Observatories between 1987 and 2012. Gray and black symbols denote individual unbinned RV measurements while orange, green, and cyan circles are annual averages for Lick, Keck, and European Southern Observatories, respectively. Error bars are displayed for annual averages only.}
\label{allthreervs}
\end{figure}

\begin{figure}
\includegraphics[scale=0.8]{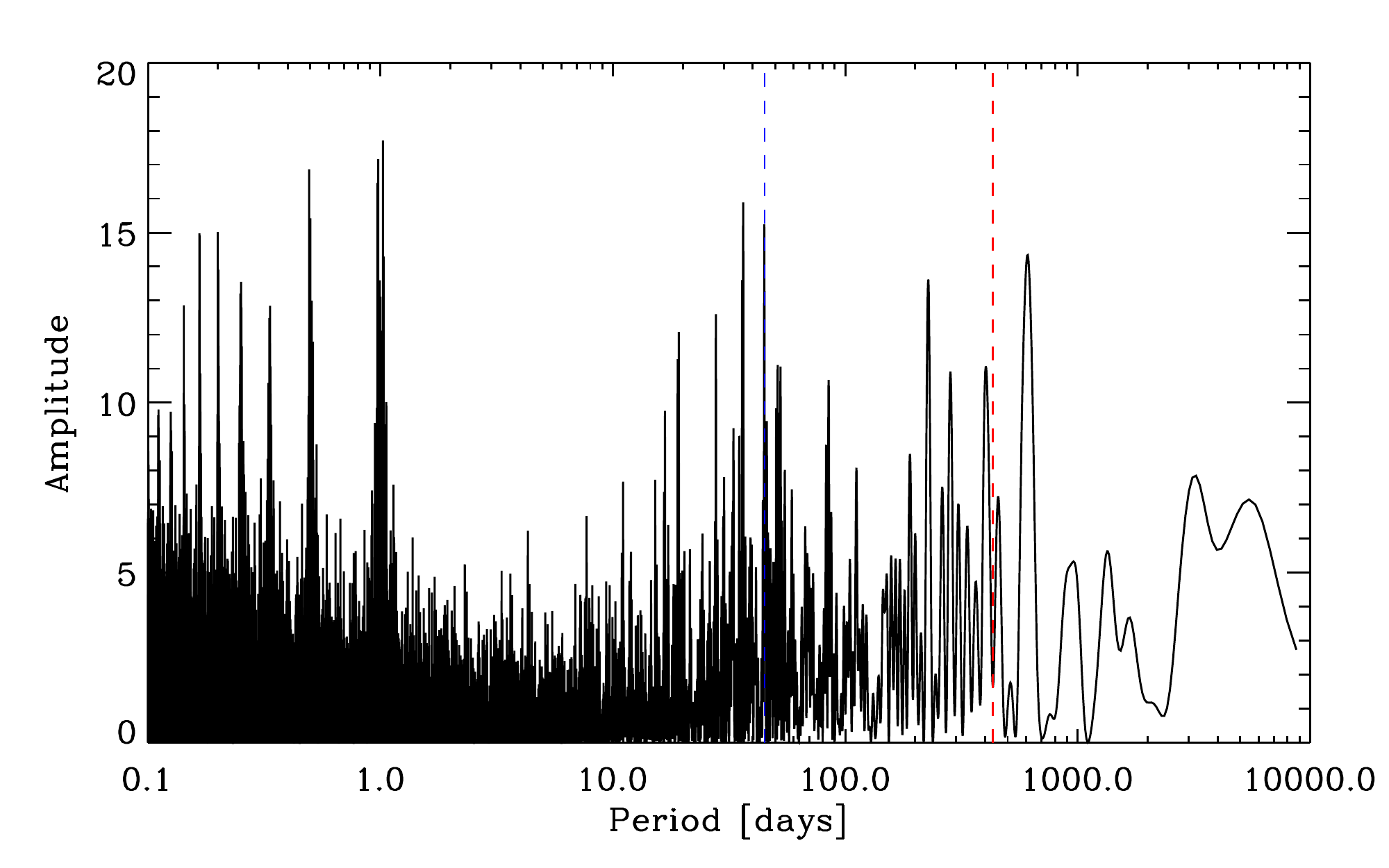}
\caption{Keck and VLT RV periodogram. Red and blue dashed lines are shown at 430 days and 45 days, respectively. The 430-day periodicity is clearly ruled out. The 45-day periodicity is not supported by the Keck RVs as shown in Fig.\ \ref{periodogram}.}
\label{vltkeck_periodogram}
\end{figure}

We use 226 publicly available RVs of Barnard's Star spanning 6 years obtained with the Very Large Telescope (VLT) to provide an independent assessment of periodicities, or lack thereof \citep{Zechmeister2009}. The observations were made using an iodine cell to precisely calibrate the wavelength scale and model the instrumental profile \citep{Zechmeister2009}. We bin the VLT RVs in intervals of two hours and add a jitter of 2 \ms{} in quadrature to the reported uncertainties for consistency with the treatment of the Keck RVs and to slightly soften the relative weights of the RVs. We also adjust the VLT RVs to have the same zero-point as the Keck RVs. Fig.\ \ref{allthreervs} shows the RVs from the VLT, with the individual VLT measurements shown in gray crosses and the annual averages in cyan circles. The RMS scatter of the VLT RVs (3.4 \ms{}) is slightly larger than the RMS of our Keck RVs (2.5 \ms{}). We construct a periodogram by combining both post-upgrade Keck and full VLT 
data as shown in Fig.\ \ref{vltkeck_periodogram}. \cite{Zechmeister2009} reported a periodicity at 45 days, which they attributed to stellar activity. This prospective period does not appear in our RVs (see Fig.\ \ref{periodogram}), thereby providing further confirmation that this periodicity is not likely to be due to a planetary companion. However, it should be noted that the VLT data and the Keck post-upgrade data have only a relatively short temporal overlap of $\sim2$ years. The periodogram of combined set of RVs from Keck and VLT shown in Fig.\ \ref{vltkeck_periodogram} reveals that the peak at 430 days is no longer prominent, implying that the VLT RVs do not support this period, thus providing additional support to rule out this periodicity.

There is a possible periodicity of $\sim7$ years seen in the Keck and VLT RVs in Fig.\ \ref{allthreervs} and as a small peak in the periodogram of those RVs in Fig.\ \ref{vltkeck_periodogram}. We combine the full Keck (both pre- and post-upgrade) and VLT RV data sets to carry out a Keplerian fit, excluding the Lick RVs due to their large errors. Although the pre-upgrade data have slightly larger measurement errors compared to the post-upgrade data, we fit all of the Keck RVs spanning 15 years because we are searching for a long periodicity of $\sim7$ years. We limit the eccentricity to 0.8 and provide an initial guess for the period of 2500 days. We also carry out a Keplerian fit with the same initial conditions using individual data sets to ensure that these fits yield consistent results. The best fit results are shown in Table\ 4. The result depends drastically on the data sets we use, casting doubt on the reality of any planet near this period. Both Keck and VLT RVs are sy
stematically high by $\sim3$ \ms{} during 2001, and the Keplerian fit cleverly places the next periastron outside our observing window, during 2010 and 2011, where we have no RV observations. The probability that a periastron passage would occur just where there are no data leaves us suspicious of a spurious fit. The periodogram in Fig.\ \ref{vltkeck_periodogram} also does not support this period.

Next, we further exploit the periodogram analysis method to determine the masses of planets
which could have been detected as a function of period. For each
orbital period, we construct synthetic velocity curves for planets in
circular orbits with a variety of values of minimum planet mass
\msini{}. Each synthetic velocity curve is sampled at the actual times of observation, thereby
preserving the window function. To each synthetic velocity in the
set, we add ``noise'' derived from the scrambled
velocities themselves. This approach clearly overestimates the actual velocity
errors, because the observed velocities may contain a genuine low
level signal.

For each value of orbital period and planet mass ($P$, \msini{}) we
construct 100 synthetic data sets consisting of a fixed-phase Keplerian signal plus ``noise,'' derived from a new scrambling of velocities each time. A periodogram is generated for each of these realizations and compared with the original, unscrambled periodogram of post-upgrade Keck RVs. If the amplitude at the injected period in the scrambled periodogram is larger than the peak at the same period in the original, unscrambled periodogram for all 100 trials, then we rule out that planet. This stringent choice of threshold yields a slightly more conservative mass limit compared to, for instance, a threshold that requires the above criterion to be satisfied for only 90 trials. For each period, we increase the mass of a fictitious planet until we reach a mass large enough to be detected. These minimum detectable masses are then recorded and are plotted as a function of period in Fig.\ \ref{circ_detec} (solid black line). The gray line shows the detection curve corresponding to o
nly 90 out of 100 trials satisfying the criterion. We also try using a random phase for the Keplerian signal each time instead of a fixed phase but we find that the two resulting detection thresholds differ by an amount that is no larger than random fluctuations between the trials. The dotted line is an analytic solution for $K$ = 2.5 \ms{} corresponding to the RMS scatter, which is consistent with our detection threshold curve. This is not surprising since we should be able to detect signals with amplitudes that are comparable to or slightly lower than the noise level, given that we have a total of 121 post-upgrade, binned RV measurements. This figure indicates that planets of 2 \mearth{} are excluded from circular orbits having periods less than 10 days. We are also able to exclude planets with minimum masses (\msini) above 5 \mearth{} for orbital periods less than 200 days and those with 10 \mearth{} = 0.03 \mjup{} for orbital periods less than 2 years.

To constrain the masses of planets that are still possible in the habitable zone (HZ) around Barnard's Star, we compute the inner and outer edges of the HZ following the method outlined in \cite{Selsis2007}. This calculation adopts the albedo of a planet with either a thick H$_2$O or CO$_2$ atmosphere, which depends on the effective temperature of the host star \citep{Kasting1993}. Additionally, we assume 50\% cloud coverage and the theoretical ``water loss" limit of $T_{\rm surface}$ = 373 K. For Barnard's Star, the HZ is located between approximately 0.05--0.1 AU (orbital periods of 10--30 days), which appears to be devoid of planets with \msini{} $>$ 3 \mearth{}. The location of the HZ around Barnard's Star is consistent with those found previously \citep{Kurster2003, Kasting1993}. These values are only approximate as the details about the HZ are currently not very well known. \cite{Barnes2012} suggested that tidal heating may be a significant factor in habitability around
 M dwarfs, which demonstrates the true complexity of the problem of habitability around other stars.

\subsection{Eccentric Orbits}
We now extend our analyses to allow for eccentric orbits. The 8-year baseline of the post-upgrade RVs from Keck with precision of 2.5 \ms{} and high observation cadence, especially during the summer of 2011, allow the detection of planets in both long- and short-period orbits.

The technique we employ is the ``bootstrap method,'' a Monte Carlo analysis in the three-dimensional parameter space of \msini{}, $P$, and orbital eccentricity, $e$. We use the scrambled RVs to represent noise---thereby assuming that there are no planets---and inject a fake planet whose properties are drawn systematically from the parameter space. The initial values of argument of periapsis ($\omega$), the time of periastron passage ($T_{\rm p}$), and systemic velocity ($\Gamma$) are always set to 0\degr{}, the first date of observation since the upgrade, and 0 \ms{}, respectively, but are allowed to float during the fitting. Orbital period and eccentricity, however, are fixed to the values in the parameter grid. The initial value of $K$, a free parameter, is calculated each time since it depends on \msini{}. If the measurement errors remain unchanged, then by definition, a Keplerian fit to this synthetic signal yields \chisq{} $\sim$ 1. To test whether or not we would have m
issed the planet, we adopt the null hypothesis momentarily---which is incorrect since we injected the planets into the signal---and carry out a test to assess this false assumption. If we are able to disprove the null hypothesis, then we know that we would not have missed the planet. But if we cannot demonstrate that the null hypothesis is false, then it implies that the planet is undetectable because we are not able to distinguish between the real signal and noise.

\begin{figure}
\includegraphics[scale=0.8]{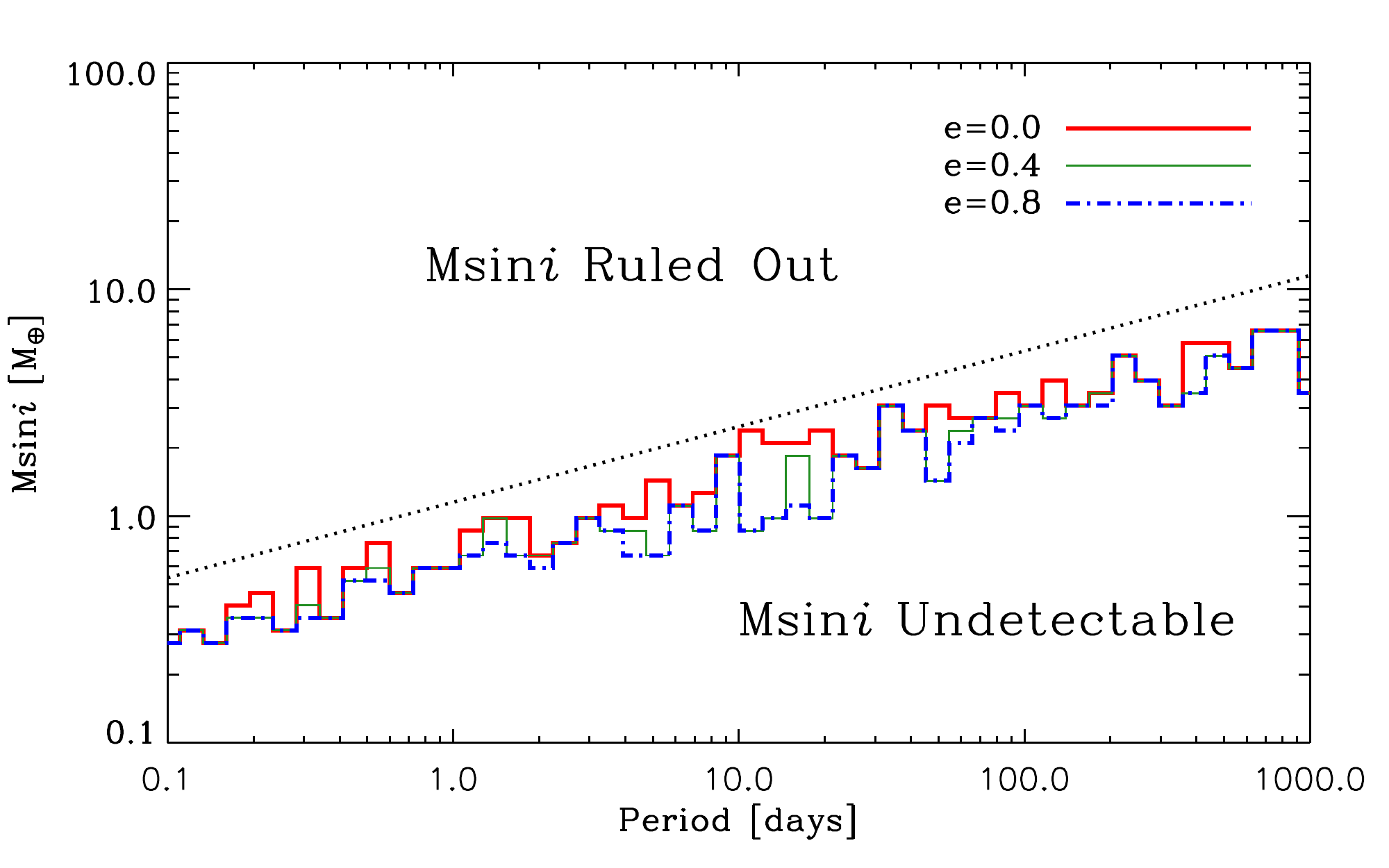}
\caption{The minimum detectable \msini{} in \mearth{} as a function of orbital period in days for different values of eccentricity, $e$, from ``bootstrap" analysis using a 90\% criterion. The dotted line is the analytic threshold corresponding to an RV semi-amplitude of $K$ = 2.5 \ms{}, the RMS noise of the RV measurements.}
\label{bootstrap_ecc_90}
\end{figure}

We first scramble the velocities and carry out a Keplerian fit---with $P$, $T_{\rm p}$, $e$, $\omega$, $K$, and $\Gamma$ as the free parameters---to obtain a \chisq{} \citep{Marcy2005}. If the original synthetic signal has a sufficiently large amplitude due to a massive planet or a close-in orbit, the RMS scatter of the scrambled velocities will be so large such that on average, the \chisq{} value will be greater than 1. We carry out 1000 such trials, allowing all parameters to float during the fitting, to produce a \chisq{} distribution and conclude that we would have detected that planet if \chisq{} of scrambled velocities is greater than \chisq{} of unscrambled velocities for at least 90 trials, corresponding to a 90\% threshold. If the scrambled \chisq{} distribution falls on or near 1, then it suggests that we would not have been able to distinguish the planet's signal from pure random noise. Fig.\ \ref{bootstrap_ecc_90} shows the detectability plot for a range of \msini
{} as a function of orbital period $P$ for different $e$. 

\begin{figure}
\includegraphics[scale=0.8]{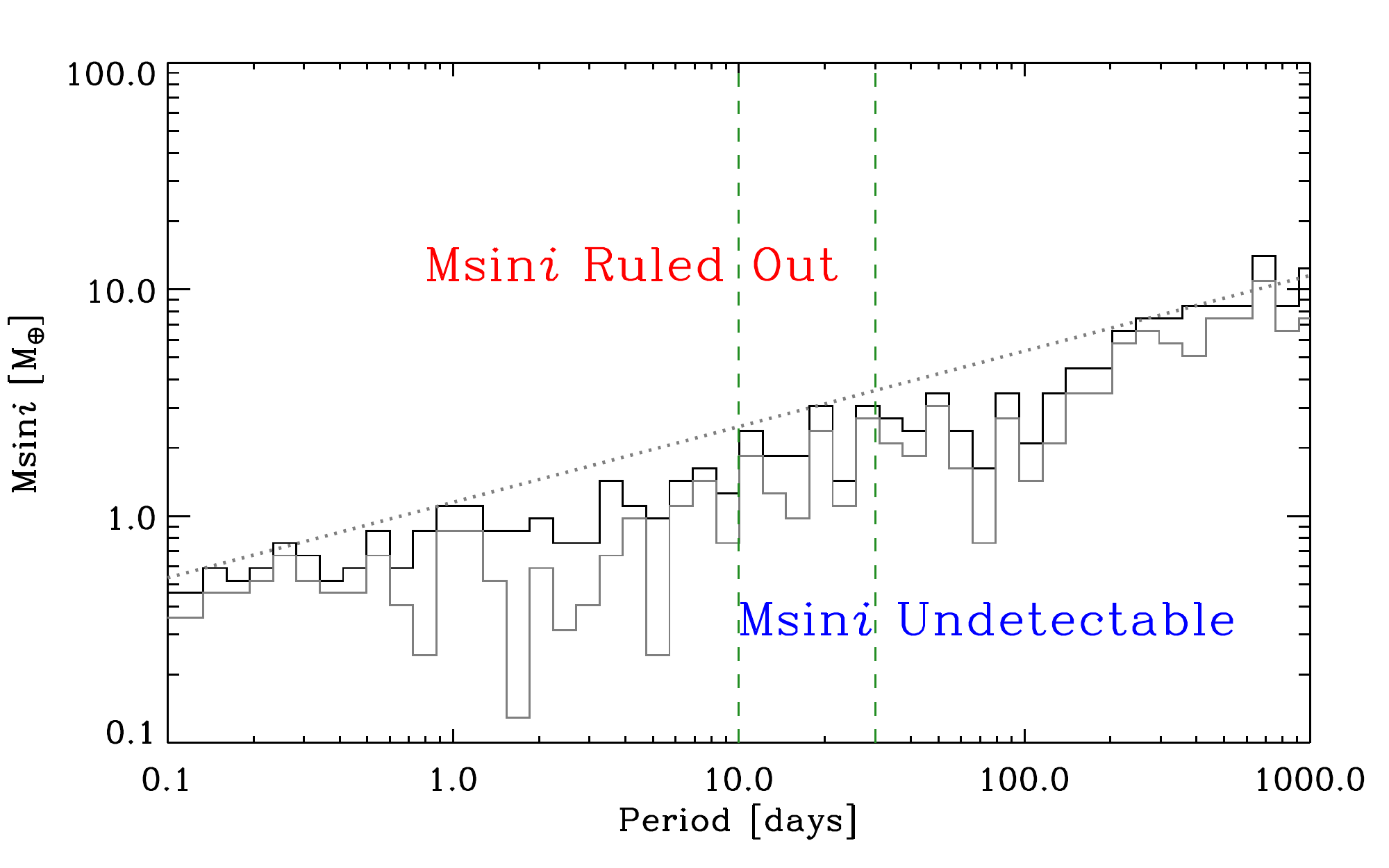}
\caption{The detection threshold in minimum mass (\msini{}) for planets as a
function of orbital period, assuming circular orbit around Barnard's Star. The dotted line is the analytic threshold corresponding to an
RV semi-amplitude of $K$ = 2.5 \ms, the RMS noise of the RV measurements. The black solid line shows a
superior detection threshold computed by 100 realizations of scrambled RVs (representing noise) at different orbital periods to determine the peak height distribution of the
periodogram from noise alone. The solid gray line corresponds to a less stringent detection criterion, which, as expected, lies below the black curve. This Monte Carlo method accounts for the
specific window function of the observations. Both the Monte Carlo and analytic methods give
similar detection thresholds. Planets with \msini{} of 1--2 \mearth{} would be
detectable for periods under 10 days, and under 5 \mearth{} for
periods under 200 days. No such planets exist around Barnard's Star,
including in its habitable zone which is shown by vertical dashed lines.}
\label{circ_detec}
\end{figure}

\begin{figure}
\includegraphics[scale=0.8]{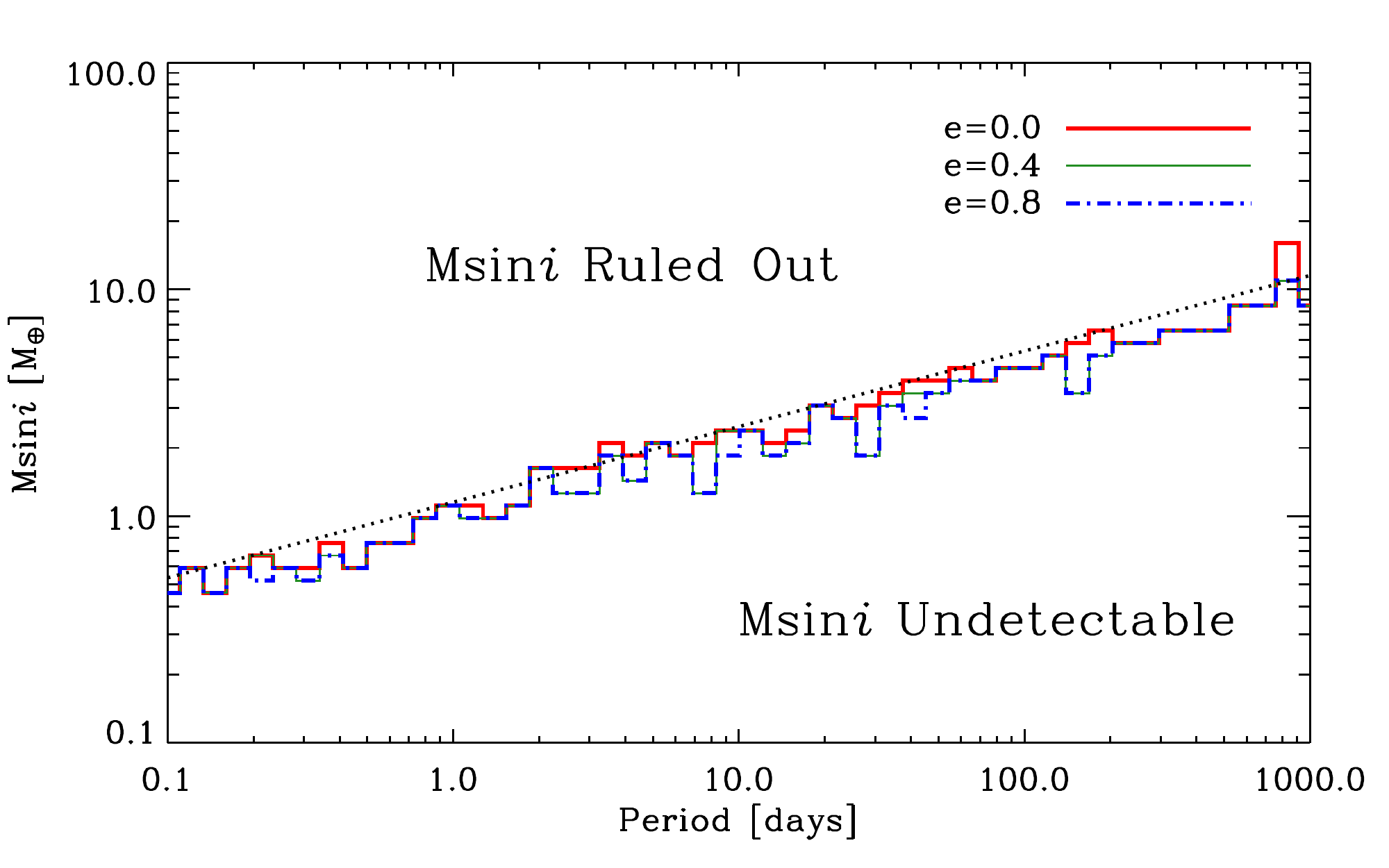}
\caption{The same as Fig.\ \ref{bootstrap_ecc_90}, but constructed using a 100\% criterion. The detection thresholds are higher compared to those in Fig.\ \ref{bootstrap_ecc_90}, as expected.}
\label{bootstrap_ecc_100}
\end{figure}

The detection thresholds in Fig.\ \ref{bootstrap_ecc_90} may be compared to those given in Fig.\ \ref{circ_detec} for circular orbits. Fig.\ \ref{bootstrap_ecc_90} shows the detection thresholds for special case of $e$ = 0 as a thick solid red line. These thresholds are approximately 40\% lower in \msini{} than those given by the $K$ = 2.5 \ms{} dotted line and slightly lower than the thresholds for circular orbits shown in Fig.\ \ref{circ_detec}. This lower \msini{} threshold reflects the somewhat arbitrary threshold of 90\% for the $\chi^2$ distribution in this current Monte Carlo test. We repeat the analysis using a 100\% criterion, and the resulting detection thresholds are shown in Fig.\ \ref{bootstrap_ecc_100}. As expected, a more stringent criterion raises the detection threshold curve (i.e., more planets will go undetected). The reader should be alerted that such statistical detection thresholds are sensitive to the arbitrary cut-off adopted in the noise distribution,
 and these thresholds are only meant to be interpreted as estimates to a factor of a few. Indeed, the modest knowledge of the temporal distribution of both RV errors and astrophysical jitter render more sophisticated threshold determinations unreliable. 

Nonetheless, for circular orbits, Fig.\ \ref{bootstrap_ecc_90} shows that planets having \msini{} = 1--2 \mearth{} would be revealed for orbital periods less than 10 days, as found in Fig.\ \ref{circ_detec}. For orbital periods up to 100 days, planets of a few Earth masses would be detected. However, no such planets were found in the observed RVs.

For planets residing in orbits of higher eccentricity, Figs.\ \ref{bootstrap_ecc_90} and \ref{bootstrap_ecc_100} (solid green and dot-dash blue curves) shows that there is slightly improved detectability. This increased sensitivity results from the periastron passage during which the star exhibits a large reflex velocity for a given orbital period compared to a circular orbit of the same period. The intense RV observations obtained during the summer of 2011 would have revealed such periastron passages, but none was seen. Thus the RVs offer higher sensitivity to eccentric orbits. These RVs therefore rule out planets in highly eccentric orbits having minimum masses of a few Earth masses out to periods of $\sim$400 days.

\section{van de Kamp's Claimed Planets}
In 1969, Peter van de Kamp first reported detection of two jovian-mass planets in circular orbits around Barnard's Star \citep{vandeKamp1969blah}. Although significant doubt
has been cast on these planets, including some by van de Kamp himself
(1982), no study has definitively ruled them out. This work is well-suited for the investigation of these claimed planets due to the long time baseline of the RVs. We consider whether
these planets would produce a detectable signal in our data. For this
analysis we use the planetary properties and orbital elements derived by van de Kamp in his
1982 paper. As his final research publication on the star, it
represents the refinement of decades of work.

\begin{figure}
\includegraphics[scale=0.8]{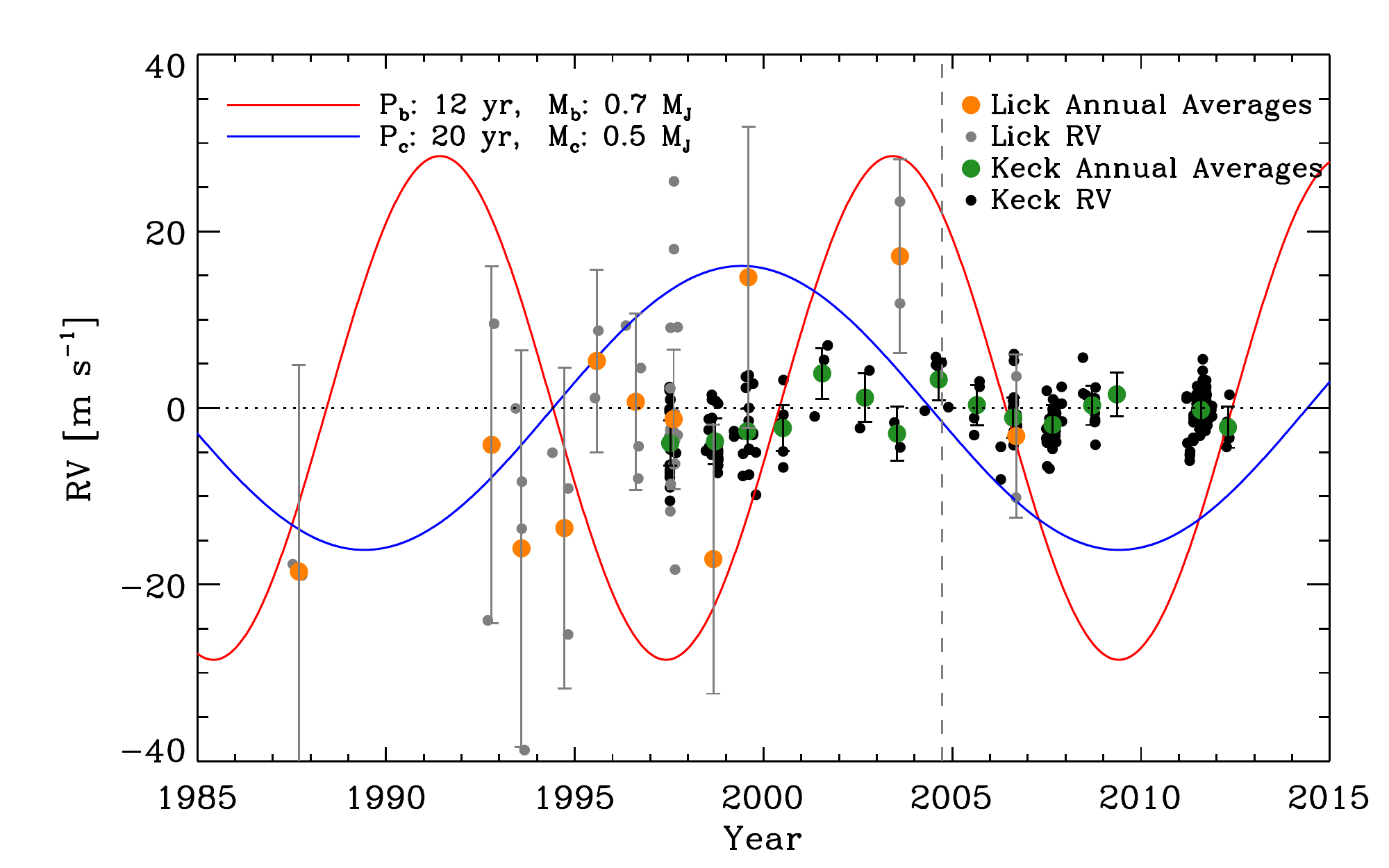}
\caption{The predicted RV curves for van de Kamp's two planets generated using the parameters from \cite{vandeKamp1982}. RV measurements from Keck and Lick Observatories are shown in black and gray for comparison. Error bars are displayed for annual averages only. The vertical dashed line denotes the CCD upgrade in 2004. Planets claimed by van de Kamp are clearly ruled out as the amplitudes of the predicted RV curves are much larger than the excursions of the observed RVs.}
\label{vdk_plot}
\end{figure}

The innermost claimed planet has mass 0.7 \mjup{} and resides in a 12-year (4383 days) circular orbit. With an inclination of 106\degr{}, the
projected mass is \msini{} = 0.672 \mjup{} = 213 \mearth{}. The second
planet in the system with mass 0.5 \mjup{} has a period of 20 years (7120 days), and lower
\msini{} = 0.45 \mjup{} = 142 \mearth{} with an inclination of 116\degr{}. We generate predicted RV signals
using these parameters (see Fig.\ \ref{vdk_plot}) and compare them
with the Keck and Lick RVs. The amplitudes of the predicted RV signals
are much larger than the RMS scatter, therefore we conclude that for the masses and inclinations found by van de Kamp, the claimed planets are clearly ruled out, by inspection. 

Even van de Kamp's earlier model consisting of a single planet with mass 1.6 \mjup{} orbiting at 4.4 AU with an inclination angle $i$ of 77\degr{} can be securely ruled out \citep{vandeKamp1963}. The corresponding $K$ for this model is 52.3 \ms{}, which would have been easily detected with our RVs. 

\subsection{Attempts to Salvage van de Kamp's Planets}
As a last resort, we investigate two ways to salvage the two planets suggested by van de Kamp. First, we consider the possibility that the RV signatures of the planets may conspire with each other to
destructively interfere during times of observation, and constructively interfere only outside of the observing windows. However, the two Keplerians would move out of phase during our 25-year observing window due to the difference in orbital periods of 8 years, and the combined signal would grow large enough to be detected. Therefore the poorly known orbital phase at the present epoch does not salvage the van de Kamp planets. Second, the true orbit of the system may be more face-on than was reported by van de Kamp ($i_1$ = 106\degr{}, $i_2$ = 116\degr{}). A nearly face-on orbit leaves only a very small radial component to be detected, below the detection threshold of a radial velocity search.

We adopt van de Kamp's (1982) stated uncertainties in the Thiele-Innes constants \citep{Aitken1935} to derive the uncertainty in the orbital inclination. It is interesting to note that these Thiele-Innes constants do not yield the inclination and nodes listed in \cite{vandeKamp1982}. Following \cite{Wright2009}, we obtain $i_1$ = 151\degr{}, $\Omega_1$ = 52\degr{}, $i_2$ = 140\degr{}, and $\Omega_2$ = 35\degr{}. We suspect that van de Kamp used only one significant figure when reporting the Thiele-Innes constants, and thus the three significant figures in $i$ and $\Omega$ are actually false precision. It is likely that he calculated and made figures with more significant figures, but only reported what was significant in \cite{vandeKamp1982}. We calculate the distribution of posterior orbital inclinations and find that the representative inclination for both planets is $137\pm28\degr$, with an asymmetric tail that admits inclinations larger than 180\degr. This value for $i$ i
s approximate because van de Kamp only provided a single significant figure on his masses. In principle, one should calculate a separate $i$ for each planet. Nevertheless, our calculation implies that face-on orbits are indeed marginally consistent with his solution. 

We perform a thorough search for 2-planet orbital solutions
consistent with van de Kamp's claims. We build a 100$\times$100 grid of
orbital periods for the two planets with periods uniformly spaced
between 11--13 and 18--22 years. At each grid point we use the RVLIN \citep{Wright2009} package to find the best fit orbital solution, iterating the fit six times. To ensure that the fitter converges on the best fit solution, we try three starting values of $T_{\rm p}$ randomly selected from a uniform distribution within $P/2$ of the median date of observation, each paired with initial values of $e$
randomly selected from a uniform distribution between 0--0.7.  We
run each of these three fits with and without floating periods. We take the best fit solution from these three fits and record the RV amplitudes of the two planets at each grid point in both the fixed- and free-period cases. In all cases we restrict $e <0.8$.

We find that the range of returned RV semi-amplitudes vary from 0--17 \ms{},
implying that in no case is $K >17$ \ms{} warranted by the
data.

\begin{figure}
\includegraphics[scale=0.8]{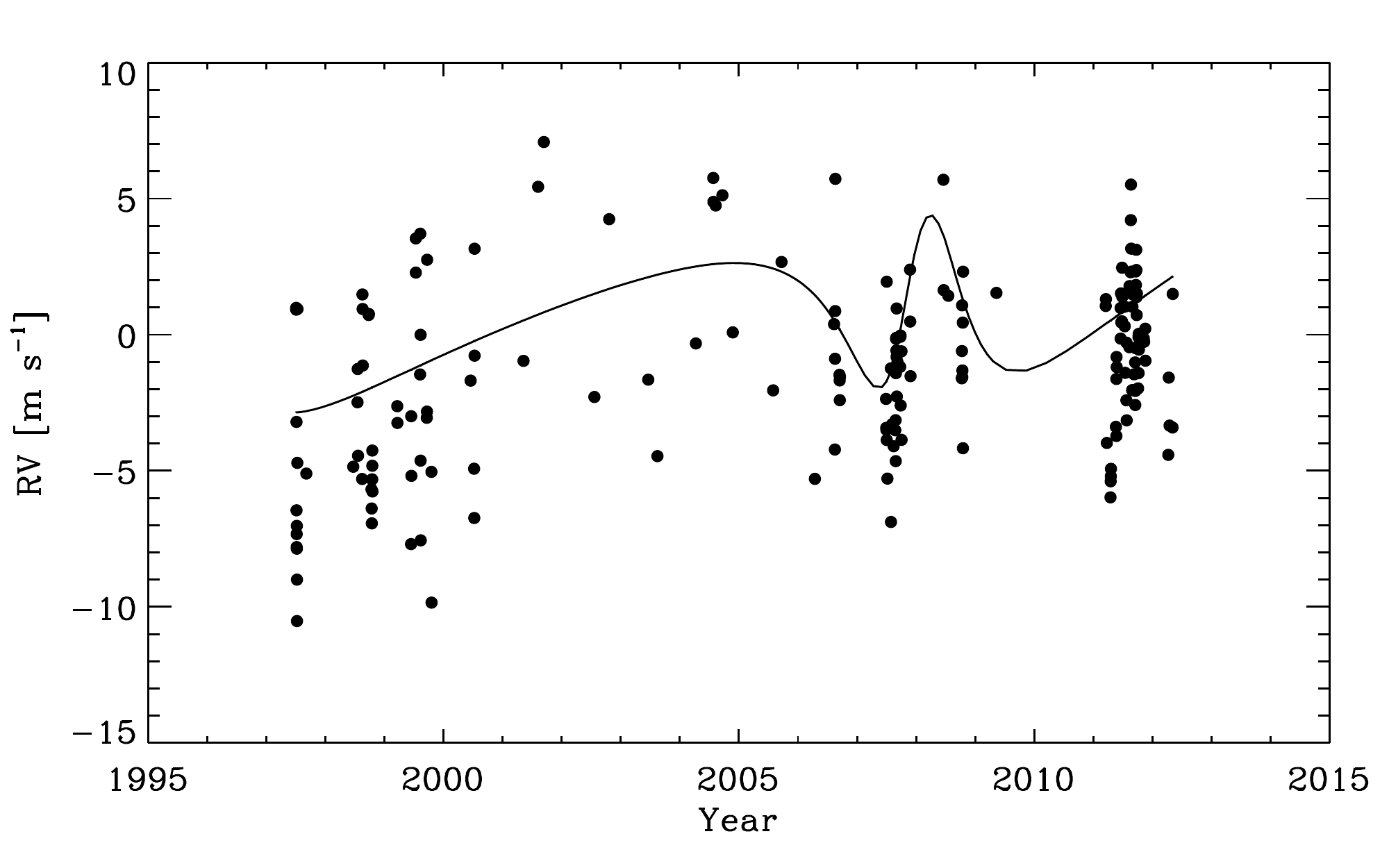}
\caption{The RV curve produced using the highest-\msini{} best fit orbital solution from the RVLIN 2-planet fit. This solution corresponds to \msini{} = 0.22 and 0.23 \mjup{} for the inner and outer planets, respectively. The value of $i$ required for this solution to be consistent with van de Kamp's astrometry is $\gtrsim160$\degr{}. Black circles are pre- and post-upgrade Keck RVs.}
\label{rvlin}
\end{figure}

The orbital solution from the fit yielding the largest planetary masses (\msini{} of 0.23 and 0.22 \mjup{}) is shown in Fig.\ \ref{rvlin}. In order for such solutions to be consistent with van de Kamp's astrometry, the value of $i$ would need to be $\gtrsim160$\degr{}. We compute the probability that the true $\sin i$ is lower than $\sin i_{\rm crit}$, assuming random orientation of the orbit in space:

\begin{equation}
{\pr}(\sin i < \sin i_{\rm crit}) = 1 - |\cos{i_{\rm crit}}|\;\;.
\end{equation}

\begin{figure}
\includegraphics[scale=0.8]{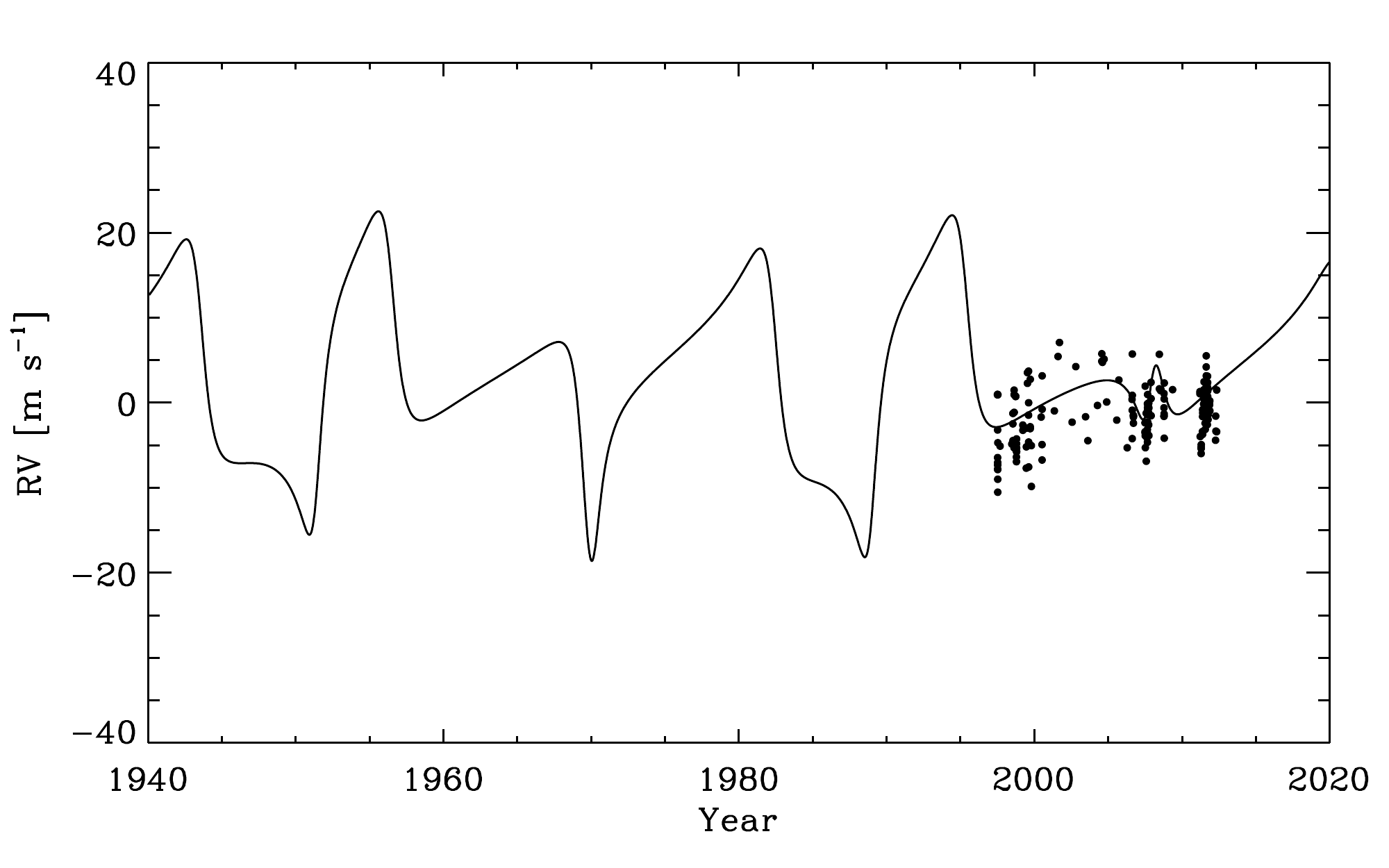}
\caption{The full RV history from 1940 to today generated using the best fit orbital solution from the RVLIN 2-planet fit. Black circles denote pre- and post-upgrade Keck RVs. It is possible that the RV signatures of van de Kamp's planets destructively interfere during the times of observations, thereby avoiding detection, but this situation is extremely unlikely.}
\label{19402020}
\end{figure}

This calculation is done a priori, and ignores the likely masses of planets orbiting M dwarfs or any knowledge of the star's astrometry and RVs. The probability that the true $\sin i$ is less than $\sin160$\degr{} or $\sin20$\degr{} is 6.0\%. This example is rather contrived, as it would require both van de Kamp to have had underestimated, not overestimated, the value of $i$, and our observations to have been timed ``just so" as to avoid the largest radial velocity excursions. Fig.\ \ref{19402020} shows the full RV history going back to the times of van de Kamp's observations, and demonstrates that while it is possible that the van de Kamp planets could ``hide" in this manner, it would require a rather insidious conspiracy of nature.

An alternative calculation is to estimate the inclination angles $i_1$ and $i_2$ that would sufficiently reduce the amplitude of the signal such that the predicted signals become consistent with the scatter in the RVs. Our calculations indicate that planets with inclination angles greater than $i_{1, \rm crit}$ = 11\degr{} (or $<$ 169\degr{}) and $i_{2, \rm crit}$ = 19\degr{} (or $<$ 161\degr{}) would be detectable. The probabilities that the orbit is more face-on than $i_{1, \rm crit}$ = 11\degr{} or 169\degr{} and $i_{2, \rm crit}$ = 19\degr{} or 161\degr{} are therefore 1.8\% and 5.4\%, respectively. Thus, the likelihood that the claimed planets are undetectable due to an error in the reported inclination angles is extremely low.

\section{Conclusion}
We have established firm upper limits to the minimum masses (\msini{}) of
planets around Barnard's Star for orbital periods ranging from a few
hours to 20 years. For orbital periods under 10 days, planets with \msini{} greater than two Earth masses would have been detected, but were not seen. For orbital periods under 100 days,
planets with minimum masses under $\sim3$ \mearth{} would have
been detected, but none was found. For periods under 2 years, planets
with minimum masses over 10 \mearth{} are similarly ruled out.

The two planets claimed by Peter van de Kamp are extremely unlikely by these
25 years of precise RVs. We frankly pursued this quarter-century
program of precise RVs for Barnard's Star with the goal of examining
anew the existence of these historic planets. Indeed, Peter van de
Kamp remains one of the most respected astrometrists of all time for
his observational care, persistence, and ingenuity. But there can be little doubt now that van de Kamp's two putative planets do not exist.

Even van de Kamp's model of a single-planet having 1.6 \mjup{} orbiting at 4.4 AU \citep{vandeKamp1963} can be securely ruled out. The RVs from the Lick and
Keck Observatories that impose limits on the stellar reflex velocity
of only a few meters per second simply leave no possibility of
Jupiter-mass planets within 5 AU, save for unlikely face-on orbits.

The lack of planets above a few Earth masses near Barnard's Star runs counter to the
discoveries of numerous mini-Neptunes, with sizes and masses slightly
above those of Earth, found recently around M dwarfs. A detailed
analysis of the planet candidates from the NASA Kepler mission shows
an increasing number of small planets (2--4 R$_{\oplus}$) around
stars of decreasing mass, including the M dwarfs \citep{Howard2011}.

\cite{Howard2011} determined occurrence rates for planets with orbital periods
less than 50 days. For planets of 2--4 R$_{\oplus}$, the occurrence is 10\% for G-type stars.
But the occurrence of such low-mass planets linearly increases with
decreasing $T_{\rm eff}$, reaching seven times more abundant around cool stars
(3600--4100 K) than around the hottest stars in the Kepler sample (6600--7100 K).
Thus Kepler finds a large occurrence of 2--4 R$_{\oplus}$ planets close-in to M
dwarfs, just where our RV survey of Barnard's Star is most sensitive
to Earth-mass planets. Yet, we found no planetary companions around Barnard's Star.

Similarly, the HARPS survey for M dwarfs has revealed numerous planets with \msini{} of a few Earth masses around M dwarfs \citep{Bonfils2011}. They examined 102 M dwarfs and found nine
``super-Earths," with two within the habitable zones of Gliese 581 and
Gliese 667C. Extrapolating, they found that the occurrence of ``super-Earths" in the habitable zone is $\sim41$\% for M dwarfs.

Thus, we have a lovely moment in science.  Two completely different
planet-hunting techniques, Doppler measurements by HARPS to detect the
reflex motion of stars, and brightness measurements by Kepler to
detect the transits of planets, give similar and extraordinary results. Small planets, slightly larger or more
massive than Earth, are apparently common around M dwarfs.

In contrast, the non-detection of planets above a few Earth masses around Barnard's Star remains remarkable as the detection limits here are as tight or tighter than was possible for the Kepler and HARPS surveys. The lack of planetary companions around Barnard's Star is interesting because of its low metallicity. This non-detection of nearly Earth-mass planets around Barnard's Star is surely unfortunate, as its distance of only 1.8 parsecs would render any Earth-size planets valuable targets for imaging and spectroscopy, as well as compelling destinations for robotic probes by the end of the century.

\section{Acknowledgements}

This work made use of the Exoplanet Orbit Database and the Exoplanet Data Explorer at www.exoplanets.org. We wish to thank our referee, Martin K\"{u}rster, for his valuable suggestions, and Jennifer Bartlett and Joshua Burkart for much helpful input. We acknowledge support by NASA grants NAG5-8299, NNX11AK04A, and NSF grant AST95-20443 (to GWM), and by Sun Microsystems. We also thank the Watson and Marilyn Albert SETI Chair fund, and Howard and Astrid Preston for generous donations making this research possible. This research has made use of the NASA Exoplanet Archive, which is operated by the California Institute of Technology, under contract with the National Aeronautics and Space Administration under the Exoplanet Exploration Program; NASA's Astrophysics Data System, and the SIMBAD database, operated at CDS, Strasbourg, France; and the Exoplanet Orbit Database and the Exoplanet Data Explorer at exoplanets.org. The Center for Exoplanets and Habitable Worlds is supported by 
the Pennsylvania State University, the Eberly College of Science, and the Pennsylvania Space Grant Consortium. We thank R.Paul Butler and Steven Vogt for help making observations. We thank the W.M Keck Observatory and Lick Observatory for their generous allocations of telescope time, and we thank the State of California for its support of operations at both observatories. We thank the W.M. Keck Foundation and NASA for support that made the Keck Observatory possible. We thank the indigenous Hawaiian people for the use of their sacred mountain, Mauna Kea.

\bibliographystyle{apj}
\bibliography{barnard_bib.bib}

\clearpage

\begin{deluxetable}{cc}
\tablecaption{Barnard's Star's Parameters}
\tablenum{1}

\tablehead{\colhead{Parameter} & \colhead{Barnard's Star}}

\startdata
$R$ (\rsun) & $0.199\pm0.006$ \tablenotemark{1,2,3}\\
$M$ (\msun) & $0.158\pm0.013$ \tablenotemark{1,4}\\
$L$ ($\rm{L_{\odot}}$) & $(3.46\pm0.17)\times10^{-3}$ \tablenotemark{5}\\
$V_{\rm mag}$ & 9.511 \tablenotemark{6}\\
$T_{\rm eff}$ (K) & $3134\pm102$ \tablenotemark{5}\\
$V\,\sin i$ (\kms) & $<2.5$ \tablenotemark{7}\\
$d$ (pc) & $1.824\pm0.005$ \tablenotemark{8}\\
$\mu$ (arcsec yr$^{-1}$)& $10.3700\pm0.0003$ \tablenotemark{9}\\
$\pi$ (arcsec) & $0.5454\pm0.0003$ \tablenotemark{9}
\enddata
\label{table:properties}
\tablenotetext{1}{\cite{Muirhead2012b}}
\tablenotetext{2}{\cite{Lane2001}}
\tablenotetext{3}{\cite{Segransan2003}}
\tablenotetext{4}{\cite{Delfosse2000}}
\tablenotetext{5}{\cite{Dawson2004}}
\tablenotetext{6}{\cite{Koen2010}}
\tablenotetext{7}{\cite{Browning2010}}
\tablenotetext{8}{\cite{Cutri2003}}
\tablenotetext{9}{\cite{Benedict1999}}
\end{deluxetable}
 
\clearpage
 
\LongTables

\begin{deluxetable}{cccc}
\tablecaption{Keck Radial Velocities for Barnard's Star}
\tablenum{2}

\tablehead{
\multicolumn{1}{c}{UT date} &
\multicolumn{1}{c}{BJD} &
\multicolumn{1}{c}{RV} &
\multicolumn{1}{c}{$\sigma$}  \\
\multicolumn{1}{c}{} &
\multicolumn{1}{c}{(-2450000)} &
\multicolumn{1}{c}{(\ms)} &
\multicolumn{1}{c}{(\ms)} 
}
%\colhead{UT date} & \colhead{BJD} & \colhead{RV} & \colhead{$\sigma$} \\ 
%\colhead{} & \colhead{(-2450000)} & \colhead{(m s$^{-1}$)} & \colhead{(m s$^{-1}$)} } 

\startdata
1997/6/2 & 602.008 & 1.99 & 2.64\\
1997/6/2 & 602.016 & -0.13 & 2.70\\
1997/6/3 & 602.952 & 2.36 & 2.61\\
1997/6/3 & 602.961 & -0.40 & 2.57\\
1997/6/4 & 604.002 & -6.45 & 2.67\\
1997/6/5 & 604.836 & -3.21 & 2.59\\
1997/6/5 & 605.101 & -7.81 & 2.68\\
1997/6/6 & 605.854 & -7.33 & 2.66\\
1997/6/6 & 606.089 & -7.86 & 2.62\\
1997/6/7 & 606.896 & -7.03 & 2.56\\
1997/6/7 & 607.098 & -9.00 & 2.77\\
1997/6/8 & 607.861 & -10.52 & 2.68\\
1997/6/9 & 609.021 & -4.71 & 2.73\\
1997/6/10 & 610.111 & 0.99 & 2.86\\
1997/6/10 & 610.119 & 0.89 & 2.86\\
1997/8/5 & 665.835 & -5.10 & 2.71\\
1998/5/21 & 954.996 & -4.85 & 2.62\\
1998/6/17 & 981.874 & -2.48 & 2.68\\
1998/6/18 & 982.886 & -1.26 & 2.60\\
1998/6/19 & 984.024 & -4.45 & 2.65\\
1998/7/15 & 1009.865 & -5.30 & 2.68\\
1998/7/17 & 1011.842 & 1.48 & 2.67\\
1998/7/18 & 1012.823 & 0.94 & 2.81\\
1998/7/19 & 1013.824 & -1.13 & 2.71\\
1998/8/25 & 1050.809 & 0.72 & 2.68\\
1998/8/26 & 1051.822 & 0.76 & 2.80\\
1998/9/12 & 1068.799 & -5.68 & 2.62\\
1998/9/13 & 1069.814 & -6.39 & 2.70\\
1998/9/14 & 1070.824 & -6.50 & 2.63\\
1998/9/14 & 1070.832 & -7.35 & 2.62\\
1998/9/15 & 1071.796 & -4.20 & 2.58\\
1998/9/15 & 1071.804 & -6.49 & 2.60\\
1998/9/16 & 1072.786 & -5.51 & 2.62\\
1998/9/16 & 1072.794 & -5.12 & 2.62\\
1998/9/17 & 1073.789 & -6.00 & 2.70\\
1998/9/17 & 1073.797 & -6.52 & 2.63\\
1998/9/17 & 1073.871 & 0.48 & 2.76\\
1998/9/17 & 1073.878 & -4.82 & 2.82\\
1998/9/18 & 1074.775 & -5.76 & 2.63\\
1999/2/18 & 1228.155 & -2.63 & 2.69\\
1999/2/19 & 1229.144 & -3.24 & 2.85\\
1999/5/12 & 1311.037 & -7.70 & 2.58\\
1999/5/13 & 1312.014 & -3.00 & 2.72\\
1999/5/14 & 1313.040 & -5.19 & 2.70\\
1999/6/11 & 1341.021 & 2.28 & 2.63\\
1999/6/12 & 1341.901 & 3.54 & 2.77\\
1999/7/8 & 1367.849 & -1.46 & 2.78\\
1999/7/9 & 1368.839 & 3.71 & 2.71\\
1999/7/10 & 1370.007 & -0.01 & 2.81\\
1999/7/11 & 1370.909 & -4.63 & 2.64\\
1999/7/12 & 1371.905 & -7.56 & 2.78\\
1999/8/19 & 1409.854 & -3.05 & 2.71\\
1999/8/20 & 1410.834 & -2.83 & 2.68\\
1999/8/21 & 1411.842 & 2.75 & 2.67\\
1999/9/17 & 1438.734 & -5.04 & 2.67\\
1999/9/18 & 1439.727 & -9.85 & 2.62\\
2000/5/15 & 1680.066 & -1.69 & 2.71\\
2000/6/7 & 1702.982 & -4.92 & 2.76\\
2000/6/8 & 1703.970 & -6.74 & 2.62\\
2000/6/9 & 1705.015 & 3.16 & 2.78\\
2000/6/10 & 1705.947 & -0.77 & 2.66\\
2001/4/7 & 2007.122 & -0.96 & 2.95\\
2001/7/7 & 2097.941 & 5.43 & 2.87\\
2001/8/12 & 2133.768 & 7.08 & 2.91\\
2002/6/20 & 2445.924 & -2.29 & 2.92\\
2002/9/20 & 2537.794 & 4.24 & 2.76\\
2003/5/17 & 2777.116 & -1.65 & 3.06\\
2003/7/14 & 2834.852 & -4.46 & 3.41\\
2004/3/7 & 3072.139 & -0.32 & 3.00\\
2004/6/24 & 3180.896 & 5.76 & 3.07\\
2004/6/25 & 3181.890 & 4.88 & 3.23\\
2004/7/9 & 3195.827 & 4.75 & 2.99\\
2004/8/20 & 3237.891 & 5.12 & 2.38\\
2004/10/23 & 3301.742 & 0.08 & 2.40\\
2005/6/29 & 3550.910 & -3.08 & 2.35\\
2005/6/29 & 3550.917 & -1.14 & 2.31\\
2005/8/20 & 3602.849 & 2.40 & 2.28\\
2005/8/20 & 3602.855 & 3.00 & 2.34\\
2006/3/12 & 3807.119 & -8.10 & 2.56\\
2006/3/12 & 3807.125 & -4.41 & 2.36\\
2006/3/12 & 3807.130 & -4.39 & 2.38\\
2006/7/10 & 3926.955 & 0.04 & 2.33\\
2006/7/10 & 3926.962 & 0.82 & 2.41\\
2006/7/15 & 3931.895 & -4.22 & 2.41\\
2006/7/16 & 3932.852 & -2.64 & 2.36\\
2006/7/16 & 3932.858 & 0.75 & 2.34\\
2006/7/17 & 3933.848 & 0.46 & 2.35\\
2006/7/17 & 3933.854 & 1.19 & 2.28\\
2006/7/18 & 3934.816 & 6.09 & 2.35\\
2006/7/18 & 3934.822 & 5.34 & 2.36\\
2006/8/13 & 3960.895 & -2.18 & 2.32\\
2006/8/13 & 3960.902 & -0.76 & 2.32\\
2006/8/14 & 3961.856 & -1.35 & 2.29\\
2006/8/14 & 3961.863 & -2.04 & 2.31\\
2006/8/15 & 3962.785 & -2.10 & 2.37\\
2006/8/15 & 3962.791 & -2.64 & 2.29\\
2006/8/16 & 3963.830 & -1.19 & 2.35\\
2006/8/16 & 3963.835 & -1.97 & 2.41\\
2007/5/26 & 4247.035 & -2.36 & 2.35\\
2007/5/27 & 4248.126 & -3.42 & 2.54\\
2007/5/29 & 4249.970 & -3.50 & 2.34\\
2007/5/30 & 4251.012 & 1.95 & 2.30\\
2007/5/31 & 4251.974 & -3.87 & 2.32\\
2007/6/4 & 4256.002 & -3.95 & 2.26\\
2007/6/4 & 4256.008 & -6.61 & 2.26\\
2007/6/26 & 4277.915 & -1.24 & 2.47\\
2007/6/27 & 4278.971 & -6.88 & 2.46\\
2007/7/4 & 4285.972 & -3.13 & 2.58\\
2007/7/4 & 4285.979 & -3.42 & 2.49\\
2007/7/13 & 4294.988 & -4.10 & 2.34\\
2007/7/23 & 4304.961 & -3.51 & 2.46\\
2007/7/24 & 4305.962 & -1.29 & 2.36\\
2007/7/25 & 4306.939 & -3.14 & 2.47\\
2007/7/26 & 4307.989 & -4.65 & 2.37\\
2007/7/27 & 4308.958 & -1.41 & 2.43\\
2007/7/28 & 4309.954 & -0.15 & 2.31\\
2007/7/29 & 4310.947 & -0.12 & 2.25\\
2007/7/30 & 4311.945 & -0.58 & 2.27\\
2007/7/31 & 4312.940 & -0.81 & 2.33\\
2007/8/1 & 4313.937 & 0.96 & 2.35\\
2007/8/2 & 4314.978 & -2.27 & 2.28\\
2007/8/6 & 4318.849 & -1.01 & 2.36\\
2007/8/23 & 4335.851 & 0.63 & 2.28\\
2007/8/23 & 4335.857 & -3.11 & 2.30\\
2007/8/24 & 4336.887 & -0.08 & 2.29\\
2007/8/25 & 4337.835 & -0.03 & 2.24\\
2007/8/26 & 4338.880 & -2.60 & 2.29\\
2007/8/31 & 4343.813 & -0.61 & 2.28\\
2007/9/1 & 4344.937 & -3.86 & 2.36\\
2007/10/23 & 4396.709 & 2.39 & 2.30\\
2007/10/24 & 4397.711 & 0.48 & 2.32\\
2007/10/26 & 4399.735 & -1.53 & 2.40\\
2008/5/16 & 4602.960 & 5.69 & 2.30\\
2008/5/17 & 4604.002 & 1.64 & 2.32\\
2008/6/16 & 4633.900 & 1.43 & 2.42\\
2008/9/8 & 4717.752 & -1.60 & 2.36\\
2008/9/9 & 4718.755 & -0.60 & 2.26\\
2008/9/10 & 4719.772 & 1.08 & 2.34\\
2008/9/11 & 4720.751 & -1.56 & 2.30\\
2008/9/12 & 4721.740 & -1.32 & 2.23\\
2008/9/14 & 4723.758 & 0.44 & 2.24\\
2008/9/15 & 4724.764 & -4.17 & 2.36\\
2008/9/16 & 4725.760 & 2.31 & 2.38\\
2009/4/8 & 4930.058 & 1.54 & 2.49\\
2011/2/14 & 5607.137 & 1.05 & 2.36\\
2011/2/15 & 5608.123 & 1.31 & 2.36\\
2011/2/21 & 5614.157 & -3.98 & 2.45\\
2011/3/13 & 5634.092 & -5.98 & 2.24\\
2011/3/14 & 5635.114 & -5.39 & 2.33\\
2011/3/15 & 5636.150 & -5.20 & 2.24\\
2011/3/16 & 5637.120 & -4.93 & 2.35\\
2011/4/16 & 5668.033 & -3.39 & 2.32\\
2011/4/19 & 5671.072 & -3.72 & 2.28\\
2011/4/20 & 5671.999 & -1.63 & 2.25\\
2011/4/21 & 5672.981 & -0.82 & 2.31\\
2011/4/22 & 5673.970 & -1.19 & 2.33\\
2011/5/15 & 5697.132 & -0.14 & 2.40\\
2011/5/16 & 5697.890 & 0.98 & 2.34\\
2011/5/18 & 5699.887 & 1.52 & 2.31\\
2011/5/19 & 5701.125 & 0.45 & 2.33\\
2011/5/23 & 5704.874 & 0.46 & 2.32\\
2011/5/24 & 5705.868 & 0.49 & 2.27\\
2011/5/25 & 5706.858 & 2.46 & 2.30\\
2011/5/26 & 5707.868 & 1.39 & 2.38\\
2011/6/10 & 5723.028 & 0.31 & 2.38\\
2011/6/11 & 5723.875 & 1.03 & 2.36\\
2011/6/14 & 5726.949 & -1.40 & 2.35\\
2011/6/21 & 5733.839 & -2.41 & 2.32\\
2011/6/22 & 5734.888 & -0.29 & 2.24\\
2011/6/23 & 5735.855 & -3.15 & 2.30\\
2011/7/9 & 5751.776 & -0.46 & 2.33\\
2011/7/10 & 5752.769 & 1.79 & 2.35\\
2011/7/17 & 5759.775 & 2.29 & 2.30\\
2011/7/18 & 5760.765 & 4.21 & 2.26\\
2011/7/19 & 5761.796 & 5.52 & 2.26\\
2011/7/20 & 5762.837 & 3.16 & 2.27\\
2011/7/21 & 5763.782 & 2.32 & 2.31\\
2011/7/26 & 5768.759 & -2.04 & 2.24\\
2011/7/27 & 5769.769 & 1.49 & 2.28\\
2011/7/28 & 5770.776 & 1.02 & 2.31\\
2011/8/8 & 5781.763 & -1.46 & 2.30\\
2011/8/14 & 5787.891 & -1.02 & 2.33\\
2011/8/15 & 5788.921 & -2.58 & 2.32\\
2011/8/16 & 5789.784 & -2.07 & 2.31\\
2011/8/17 & 5790.851 & -0.49 & 2.31\\
2011/8/18 & 5791.905 & 2.32 & 2.30\\
2011/8/19 & 5792.857 & 1.83 & 2.23\\
2011/8/20 & 5793.733 & -2.02 & 2.28\\
2011/8/21 & 5794.788 & 3.12 & 2.36\\
2011/8/22 & 5795.891 & 2.38 & 2.31\\
2011/8/23 & 5796.736 & 1.38 & 2.27\\
2011/8/24 & 5797.748 & 0.72 & 2.24\\
2011/8/25 & 5798.848 & 1.51 & 2.22\\
2011/9/2 & 5806.808 & -1.97 & 2.47\\
2011/9/4 & 5808.816 & -0.08 & 2.39\\
2011/9/5 & 5809.884 & -1.41 & 2.36\\
2011/9/6 & 5810.891 & 0.03 & 2.24\\
2011/9/7 & 5811.834 & -0.55 & 2.21\\
2011/9/10 & 5814.724 & -0.34 & 2.25\\
2011/10/7 & 5841.771 & -0.14 & 2.36\\
2011/10/9 & 5843.706 & -0.26 & 2.23\\
2011/10/16 & 5850.711 & 0.22 & 2.30\\
2011/10/17 & 5851.705 & -0.96 & 2.30\\
2012/3/5 & 5992.156 & -4.42 & 2.44\\
2012/3/8 & 5995.113 & -1.58 & 2.39\\
2012/3/13 & 6000.091 & -3.34 & 2.33\\
2012/4/1 & 6019.007 & -3.41 & 2.61\\
2012/4/2 & 6020.010 & 1.50 & 2.37
\enddata
\label{table:keckrv_table}
\end{deluxetable}

\newpage
 
\begin{deluxetable}{cccc}
\tablecaption{Lick Radial Velocities for Barnard's Star}
\tablenum{3}

\tablehead{\colhead{UT date} & \colhead{BJD} & \colhead{RV} & \colhead{$\sigma$} \\ 
\colhead{} & \colhead{(-2450000)} & \colhead{(m s$^{-1}$)} & \colhead{(m s$^{-1}$)} } 

\startdata
1987/6/11 & -3042.112 & -17.74 & 32.55\\
1987/9/10 & -2951.294 & -19.01 & 23.42\\
1992/8/11 & -1154.246 & -24.10 & 24.35\\
1992/10/11 & -1093.362 & 9.50 & 20.23\\
1993/5/7 & -885.001 & -0.09 & 32.22\\
1993/7/4 & -827.233 & -13.72 & 22.46\\
1993/7/6 & -825.161 & -8.39 & 24.43\\
1993/8/1 & -799.234 & -38.79 & 26.58\\
1994/4/27 & -530.011 & -5.11 & 24.16\\
1994/9/27 & -377.324 & -25.70 & 20.65\\
1994/9/28 & -376.357 & -9.16 & 18.14\\
1995/6/13 & -118.027 & 1.08 & 11.45\\
1995/7/15 & -86.155 & 8.71 & 10.36\\
1996/4/8 & 181.992 & 9.28 & 10.20\\
1996/8/5 & 300.759 & -8.04 & 10.19\\
1996/8/9 & 304.719 & -4.39 & 11.59\\
1996/8/30 & 325.741 & 4.48 & 10.01\\
1997/6/11 & 610.891 & 2.19 & 10.76\\
1997/6/11 & 610.906 & -2.53 & 19.14\\
1997/6/11 & 610.921 & -11.74 & 14.12\\
1997/6/14 & 613.787 & 9.04 & 12.60\\
1997/6/14 & 613.831 & -3.07 & 14.12\\
1997/6/15 & 614.841 & -8.66 & 8.67\\
1997/6/15 & 614.864 & -2.37 & 9.43\\
1997/7/11 & 640.812 & -2.67 & 9.64\\
1997/7/11 & 640.834 & -6.42 & 9.68\\
1997/7/16 & 645.801 & 25.62 & 8.99\\
1997/7/16 & 645.817 & 17.94 & 24.08\\
1997/7/26 & 655.787 & -6.38 & 8.87\\
1997/7/26 & 655.810 & -0.68 & 7.91\\
1997/7/27 & 656.786 & -18.35 & 8.93\\
1997/7/27 & 656.809 & -3.66 & 8.31\\
1997/8/21 & 681.700 & 9.10 & 10.06\\
1997/8/21 & 681.723 & -3.11 & 9.49\\
1998/8/1 & 1026.774 & -17.16 & 15.25\\
1999/7/5 & 1364.769 & 14.75 & 17.03\\
2003/7/11 & 2831.841 & 11.79 & 10.97\\
2003/7/11 & 2831.854 & 23.33 & 11.80\\
2006/8/8 & 3955.768 & -10.20 & 9.38\\
2006/8/11 & 3958.676 & 3.53 & 9.24
\enddata
\label{table:lickrv_table}
\end{deluxetable}

\begin{deluxetable}{cccc}
\tablecaption{Best Fit Keplerian Parameters}
\tablenum{4}

\tablehead{\colhead{Parameter} & \colhead{Keck \& VLT} & \colhead{Keck} & \colhead{VLT}}

\startdata
$P$ (days) & 3037 & 3064 & 2356 \\
$e$ & 0.46 & 0.80 & 0.35 \\
$T_{\rm p}$ (BJD) & 2454904 & 2454187 & 2451990 \\
$K$ (\ms{}) & 2.13 & 2.87 & 2.58 \\
\msini{} (\mearth{}) & 14.1 & 19.0 & 15.7 
\enddata
\label{table:8year_kep}
\end{deluxetable}

\end{document}